\documentclass[11pt]{article}
\usepackage{moriond,epsfig}

\newcommand{\GeV}{\ensuremath{\mathrm{Ge\kern -0.1em V}}}
\newcommand{\TeV}{\ensuremath{\mathrm{Te\kern -0.1em V}}}
\newcommand{\TeVcc}{\ensuremath{\,\mathrm{Te\kern -0.1em V\!/c}^2}}
\newcommand{\GeVcc}{\ensuremath{\,\mathrm{Ge\kern -0.1em V\!/c}^2}}
\newcommand{\MeVcc}{\ensuremath{\,\mathrm{Me\kern -0.1em V\!/c}^2}}
\newcommand{\GeVc}{\ensuremath{\mathrm{Ge\kern -0.1em V}\!/c}}

\newcommand{\ifb}{\ensuremath{\mathrm{fb^{-1}}}}

\newcommand{\Hi}{\ensuremath{\mathrm{H}}}
\newcommand{\W}{\ensuremath{\mathrm{W}}}
\newcommand{\Wjets}{\ensuremath{\mathrm{W+jets}}}

\newcommand{\WW}{\ensuremath{\W^+\W^-}}
\newcommand{\Z}{\ensuremath{\mathrm{Z}}}

\newcommand{\WZ}{\ensuremath{\W\Z}}

\newcommand{\Elp}{\ensuremath{\mathrm{\mathrm{e}}^{+}}}
\newcommand{\Elm}{\ensuremath{\mathrm{\mathrm{e}}^{-}}}
\newcommand{\Elpm}{\ensuremath{\mathrm{\mathrm{e}}^{\pm}}}

\newcommand{\Mp}{\ensuremath{\mu^{+}}}
\newcommand{\Mm}{\ensuremath{\mu^{-}}}

\newcommand{\Mmp}{\ensuremath{\mu^{\mp}}}

\newcommand{\Lep}{\ensuremath{\mathrm{\ell}}}

\newcommand{\ttbar}{\ensuremath{\mathrm{t}\bar{\mathrm{t}}}}

\newcommand{\mHi}{\ensuremath{m_{\mathrm{H}}}}

\newcommand{\mll}{\ensuremath{m_{\Lep\Lep}}}

\newcommand{\pt}{\ensuremath{p_\mathrm{T}}}

\newcommand{\ptlmax}{\ensuremath{p_{\mathrm{T}}^{\Lep,\mathrm{max}}}}
\newcommand{\ptlmin}{\ensuremath{p_{\mathrm{T}}^{\Lep,\mathrm{min}}}}
\newcommand{\met}{\ensuremath{\Et^{\mathrm{miss}}}}
\newcommand{\delphill}{\ensuremath{\Delta\phi_{\Lep\Lep}}}

\newcommand{\Et}{\ensuremath{E_\mathrm{T}}}

\newcommand{\delphimetll}{\ensuremath{\Delta\phi_{\met\Lep\Lep}}}
\newcommand{\GAMMA}{\ensuremath{\gamma}}
\newcommand{\mt}{\ensuremath{m_{T}}}

\newcommand{\hww}{\Hi\to\WW}

\newcommand{\wgamma}{\ensuremath{W\GAMMA}}

\newcommand{\ee}{\ensuremath{ee}}

\def\Journal#1#2#3#4{{#1} {\bf #2}, #3 (#4)}

\def\be{\begin{equation}}
\def\ee{\end{equation}}
\def\bea{\begin{eqnarray}}
\def\eea{\end{eqnarray}}

\begin{document}
\vspace*{4cm}
\title{Study of Standard Model Scalar Production in Bosonic Decay Channels in CMS}

\author{Guillelmo G\'omez-Ceballos}

\address{Massachusetts Institute of Technology, Cambridge, USA}

\maketitle\abstracts{
The status of the Standard Model Scalar Boson search in the bosonic decay channels at the 
CMS experiment at the LHC is presented. The results are based on proton-proton collisions 
data corresponding to integrated luminosities of up to 5.1 $\ifb$ at $\sqrt{s}$ = 7 $\TeV$ 
and 19.6 $\ifb$ at $\sqrt{s}$ = 8 $\TeV$. The observation of a new boson at a mass near 
126 $\GeV$ is confirmed by the analysis of the new data and first measurements of the boson 
properties are shown.}

%%%%%%%%%%%%%%%%%%%%%%%%%%%%%%%%%%%%%%%%%%%%%%%%%%%%%%%%%%%%%%%%%%%%%%%%%%%%%%%
\section{Introduction}
\label{sec:introduction}
%%%%%%%%%%%%%%%%%%%%%%%%%%%%%%%%%%%%%%%%%%%%%%%%%%%%%%%%%%%%%%%%%%%%%%%%%%%%%%%
One of the open questions in the standard model (SM) of particle
physics~\cite{SM1,SM2,SM3} is the origin of the masses of fundamental
particles. Within the SM, vector boson masses arise from the
spontaneous breaking of electroweak symmetry by the
Higgs field~\cite{Higgs1,Higgs2,Higgs3,Higgs4,Higgs5,Higgs6}. In 2012, 
the LHC experiments, ATLAS and CMS, reported the discovery of a new boson at
approximately 125~\GeV with 5 or more standard deviations each~\cite{CMSPaperCombination,AtlasPaperCombination}.
Both observations are consistent with expectations for the SM Higgs boson 
within the large statistical uncertainties.

The $\Hi \to VV$ modes have the largest sensitivity among all Higgs decays. 
In particular, the $\Hi \to \Z\Z \to 4\ell$ and $\Hi \to \gamma\gamma$ 
modes have very good mass resolution, while the $\Hi \to \W\W \to 2\ell 2\nu$ 
mode has very large signal yield. 
In these proceedings the updated analyses with the full available dataset 
by the time of the conference are summarized. The data sample 
corresponds up to 5.1 $\pm$ 0.1 $\ifb$ (19.5 $\pm$ 0.8 $\ifb$) of 
integrated luminosity collected in 2011 (2012) at a center-of-mass energy 
of 7 (8) $\TeV$ collected by the CMS experiment at the LHC. 

These proceedings are organized as follows. Section~\ref{sec:cms} briefly describes the main
components of the CMS detector used in the analyses, while Section~\ref{sec:objects} describes
the general selection used to define the objects. The following 
Sections~\ref{sec:hzz} to~\ref{sec:hzg} explain each updated final 
state analysis: $\Hi \to \Z\Z \to 4\ell$, $\Hi \to \W\W \to 2\ell 2\nu$, 
$\W\Hi \to \W\W\W \to 3\ell 3\nu$ and $\Hi \to \Z\gamma$. It is worth noting the spin 
and parity studies on $\Hi \to \Z\Z \to 4\ell$ and  $\Hi \to \W\W \to 2\ell 2\nu$ are 
summarized in the CMS Standard Model Scalar properties proceeding. By the time of 
the conference, there was no update on $\Hi \to \gamma\gamma$, and therefore the 
quoted results in Reference~\cite{CMSPaperCombination} were still valid.

All Higgs production mechanisms are considered: the gluon fusion process, the 
associated production of the Higgs boson with a $\W$ or $\Z$ boson (VH), the 
$\ttbar\Hi$ process, and the vector boson fusion (VBF) process. The SM Higgs 
boson production cross sections are taken from Reference~\cite{LHCHiggsCrossSectionWorkingGroup:2011ti}.

%%%%%%%%%%%%%%%%%%%%%%%%%%%%%%%%%%%%%%%%%%%%%%%%%%%%%%%%%%%%%%%%%%%%%%%%%%%%%%%
\section{CMS detector and event simulation}
\label{sec:cms}
%%%%%%%%%%%%%%%%%%%%%%%%%%%%%%%%%%%%%%%%%%%%%%%%%%%%%%%%%%%%%%%%%%%%%%%%%%%%%%%
The CMS detector is described in detail elsewhere~\cite{CMSdetector}. The 
key components used for this analysis are summarized here. 
A superconducting solenoid occupies the
central region of the CMS detector, providing an axial magnetic 
field of 3.8~Tesla parallel to the beam direction. Charged particle trajectories
are measured by the silicon pixel and strip tracker, which cover the
pseudo-rapidity region $|\eta| < 2.5$. Here, $\eta$ is defined as 
$\eta=-\ln{\tan{\theta/2}}$, where $\theta$ is the polar angle of the trajectory of the particle
with respect to the direction of the counterclockwise beam. A crystal electromagnetic
calorimeter (ECAL) and a brass/scintillator hadron calorimeter (HCAL)
surround the tracking volume and cover $|\eta| < 3$. A quartz-fiber
Cherenkov calorimeter (HF) extends the coverage to $|\eta| < 5$. The muon system consists of 
gas detectors embedded in the iron return yoke outside the solenoid, with a coverage to $|\eta| < 2.4$. 
The first level of the CMS trigger system, composed of custom
hardware processors, is designed to select the most interesting events
in less than 3 $\mu s$, using information from the calorimeters and muon
detectors. The High Level Trigger processor farm further
reduces the event rate to a few hundred Hz before data storage. 

Several Monte Carlo event generators are used to simulate the signal and background 
processes. For all of them, the detector response is simulated using a detailed
description of the CMS detector, based on the \textsc{geant4} 
package~\cite{Agostinelli:2002hh}. Minimum bias events are superimposed 
on the simulated events to emulate the additional pp interactions per 
bunch crossing (pile-up). These samples are re-weighted to represent the pile-up 
distribution as measured in the data. The average number of pile-up events per 
beam crossing in the 2011 data is about 10, and in the 2012 data it is about 20.

%%%%%%%%%%%%%%%%%%%%%%%%%%%%%%%%%%%%%%%%%%%%%%%%%%%%%%%%%%%%%%%%%%%%%%%%%%%%%%%
\section{Object selection}
\label{sec:objects}
%%%%%%%%%%%%%%%%%%%%%%%%%%%%%%%%%%%%%%%%%%%%%%%%%%%%%%%%%%%%%%%%%%%%%%%%%%%%%%%
The selection requirements to define the objects in the different final states 
depend on their specific characteristics, both in terms of the signal topology
and the background processes. Nevertheless, the general strategy to select 
the different objects is common for all of them, and it is described below.

Signal candidates are selected online by trigger paths requiring
the presence of one or several electrons or muons. The use of a combination of 
them make the efficiencies for events satisfying the analysis selection 
above 95\% for the final states under study.

Muon candidates are reconstructed combining two algorithms, one in which 
tracks in the silicon detector are matched to hits in the
muon system, and another in which a global fit is performed on hits
in both the silicon tracker and the muon system. Muons are 
required to be isolated to distinguish between muons 
from $\W/\Z$ boson decays and those from QCD background processes, 
which are usually in or near jets. For each muon candidate, the scalar 
sum of the transverse energy of all particles
compatible with originating from the primary vertex is reconstructed in
cones of several widths around the muon direction, excluding the
contribution from the muon itself. This information is combined using a
multivariate algorithm which exploits the differences in the differential
energy deposition between prompt muons and muons from hadron decays inside
a jet, to discriminate between signal and background.

Electron candidates are identified using a multivariate approach based on variables 
which exploit information from the tracker, the ECAL, and the combination of these two detectors. 
Electron isolation is characterized by the ratio of the sum of the
transverse energy of the particles reconstructed in a cone around the
electron, excluding the contribution from the electron itself, and the
transverse energy of the electron. Isolated electrons are selected by
requiring this ratio to be below a threshold.

For both electrons and muons corrections are applied to account for the contribution to 
the energy in the isolation cone from  the pile-up. 
A median energy density ($\rho$) is determined event by event and the pile-up contribution 
is estimated as the product of $\rho$ and an effective isolation cone area. 
This contribution is subtracted~\cite{Cacciari:subtraction} from the transverse energy in the 
isolation cone.

Hadronically decaying $\tau$ leptons are reconstructed and identified using 
an algorithm~\cite{CMS-PAS-TAU-11-001} which targets the main decay modes by selecting 
candidates with one charged hadron and up to two neutral pions, or with three charged hadrons. 

The lepton candidates are required to originate from the primary vertex of the event, which 
is chosen as the vertex with the highest $\sum \pt^2$, where the sum 
runs over all tracks associated with the vertex.

Photon candidates are reconstructed from clusters of channels in the ECAL 
around channels with significant energy  deposits,  which are merged into 
superclusters. The clustering algorithms result in almost complete recovery of 
the energy of photons in spite of the large fraction of Bremsstrahlung and converted photons. 
In the endcaps, the preshower energy is added where the preshower is present
($|\eta| > 1.65$).  The observables used in the photon selection are: isolation
variables based on the particle flow algorithm~\cite{PFT-09-001}, the
ratio of hadronic energy in the hadron calorimeter  towers behind
the supercluster to the electromagnetic energy in the supercluster, the transverse width of  the
electromagnetic shower, and an electron veto to avoid misidentifying an electron as a photon.

Jets are reconstructed using the anti-$\mathrm{k_T}$ clustering algorithm~\cite{antikt} 
with distance parameter $\Delta \mathrm{R}=0.5$, as implemented in the \textsc{fastjet} 
package~\cite{Cacciari:fastjet}.
A similar correction as for the lepton isolation is applied to account for the contribution 
to the jet energy from pile-up events. Jet energy corrections are applied as a function 
of the jet $\Et$ and $\eta$~\cite{cmsJEC}. 
%The properties of the hard jets are modified by particles from pile-up interactions. A 
%combinatorial background arises from low-$\pt$ jets from pile-up interactions which get 
%clustered into high-$\pt$ jets. In the 2012 data the number of pile-up events is larger and
%a multivariate selection is applied to separate jets from the 
%primary interaction from those reconstructed due to energy deposits associated with 
%pile-up. The discrimination is based on the differences in the jet shapes,
%in the relative multiplicity of charged and neutral components and in the different 
%fraction of transverse momentum which is carried by the hardest components.
%Within the tracker acceptance the jet tracks are also required to be compatible
%with the primary vertex. 
Events are classified according to the number of selected jets with $\Et>30~\GeV$ and $|\eta|<~4.7$. 

Neutrinos escape detection, and result in large missing transverse energy, 
$\met$, defined as the modulus of the negative 
vector sum of the transverse momenta of all reconstructed particles 
(charged or neutral) in the event~\cite{PFT-09-001}. Since the $\met$ resolution is degraded by pile-up, 
the minimum of two different observables is used. The first includes all particle candidates in the
event~\cite{PFT-09-001}. The second uses only the charged
particle candidates associated with the primary vertex. The use of both 
variables exploits the presence of a correlation between the two variables in 
signal events with genuine {\it $\met$}, and its absence otherwise, as in 
Drell-Yan events.

To suppress the top-quark background, a \textit{top tagging} technique based
on soft-muon and b-jet tagging~\cite{btagcms} is applied. 
The first method rejects events with soft muons which likely come from semileptonic b-decays coming from top-quark decays. 
The second method uses a b-jet tagging algorithm which looks for tracks with large impact parameter 
within jets. For the second method jets with $\Et>15~\GeV$ are considered. The 
rejection factor for the top-quark background is about 50\% in the 0-jet
category and above 80\% for events with at least one jet passing the selection criteria. 

%%%%%%%%%%%%%%%%%%%%%%%%%%%%%%%%%%%%%%%%%%%%%%%%%%%%%%%%%%%%%%%%%%%%%%%%%%%%%%%
\section{$\Hi \to \Z\Z \to 4\ell$ analysis}
\label{sec:hzz}
%%%%%%%%%%%%%%%%%%%%%%%%%%%%%%%%%%%%%%%%%%%%%%%%%%%%%%%%%%%%%%%%%%%%%%%%%%%%%%%
The $\Hi \to \Z\Z \to 4\ell$ analysis~\cite{CMS-PAS-HIG-13-002} presented here 
relies critically on the reconstruction, identification, and isolation of leptons.
The high lepton reconstruction efficiencies are achieved for a $\Z\Z$ system composed of two 
pairs of same-flavour and opposite-charge isolated leptons, in the measurement range 
$m_{4\ell}, m_{2\ell2\tau} > 100~\GeV$. One or both of the $\Z$ bosons can be off-shell.
%The $\Z\rightarrow 4\ell$ resonance is used in the mass range 
%$70 < m_{4\ell} < 100\GeV$ to cross-check our mass measurement method. 
The background sources include an irreducible four-lepton contribution from 
direct $\Z\Z$ (or $\Z\gamma^*$) production via $qq$ annihilation and $gg$ fusion.
Reducible contributions arise from $\Z+b\bar{b}$ and $\ttbar$, where the final states contain two isolated 
leptons and two jets producing secondary leptons. Additional background of instrumental nature arises from $\Z+jets$, 
$\Z+\gamma+jets$ and $\W\Z+jets$ events, where jets are misidentified as leptons.

A matrix element likelihood approach~\cite{CMSPaperCombination,Alwall:2007st,Avery:2012um} 
is used to construct a kinematic discriminant ($K_D$) based on the probability 
ratio of the signal and background hypotheses, 
$K_D = {\cal P_\mathrm{sig}}/({\cal P_\mathrm{sig}}+{\cal P_\mathrm{bkg}})$, 
where the likelihood ratio is defined for each value of $m_{4\ell}$, being 
$P_\mathrm{sig}$ and $P_\mathrm{bkg}$ the signal and background 
probabilities, respectively.

To improve the sensitivity to the production mechanisms, the event sample is 
split into two categories based on the jet multiplicity: events with fewer 
than two jets, and events with at least two jets. In the first category the 
transverse momentum divided by the mass of the four lepton system 
($\pt /m_{4\ell}$) is used to discriminate VBF and VH from gluon fusion. In the
second category a linear discriminant ($V_D$) is formed combining two VBF 
sensitive variables, the difference in pseudo-rapidity and the 
invariant mass of the two leading jets. The discriminant is tuned 
to separate vector boson from gluon fusion processes.

In summary, $m_{4\ell}$, $K_D$, and the two distributions after splitting into two categories 
based on the jet multiplicity are used to discriminate between signal and background. 
The reconstructed four-lepton invariant-mass distributions for the 
$4\ell$ and $2\ell2\tau$ final states are shown in Figure~\ref{fig:Mass4l} 
and compared with the expectation from SM background processes. The 
observed distribution is in good agreement with the expectation. 
The $\Z\rightarrow 4\ell$ resonance peak is 
observed with normalization and shape as expected. The measured distribution at 
higher mass is dominated by the irreducible $\Z\Z$ background. A clear peak 
around $m_{4\ell}=126$ GeV is seen.

\begin{figure}[!htb]
\vspace*{0.3cm}
\begin{center}
\begin{tabular}{cc}
\includegraphics[width=0.49\textwidth,height=0.25\textheight]{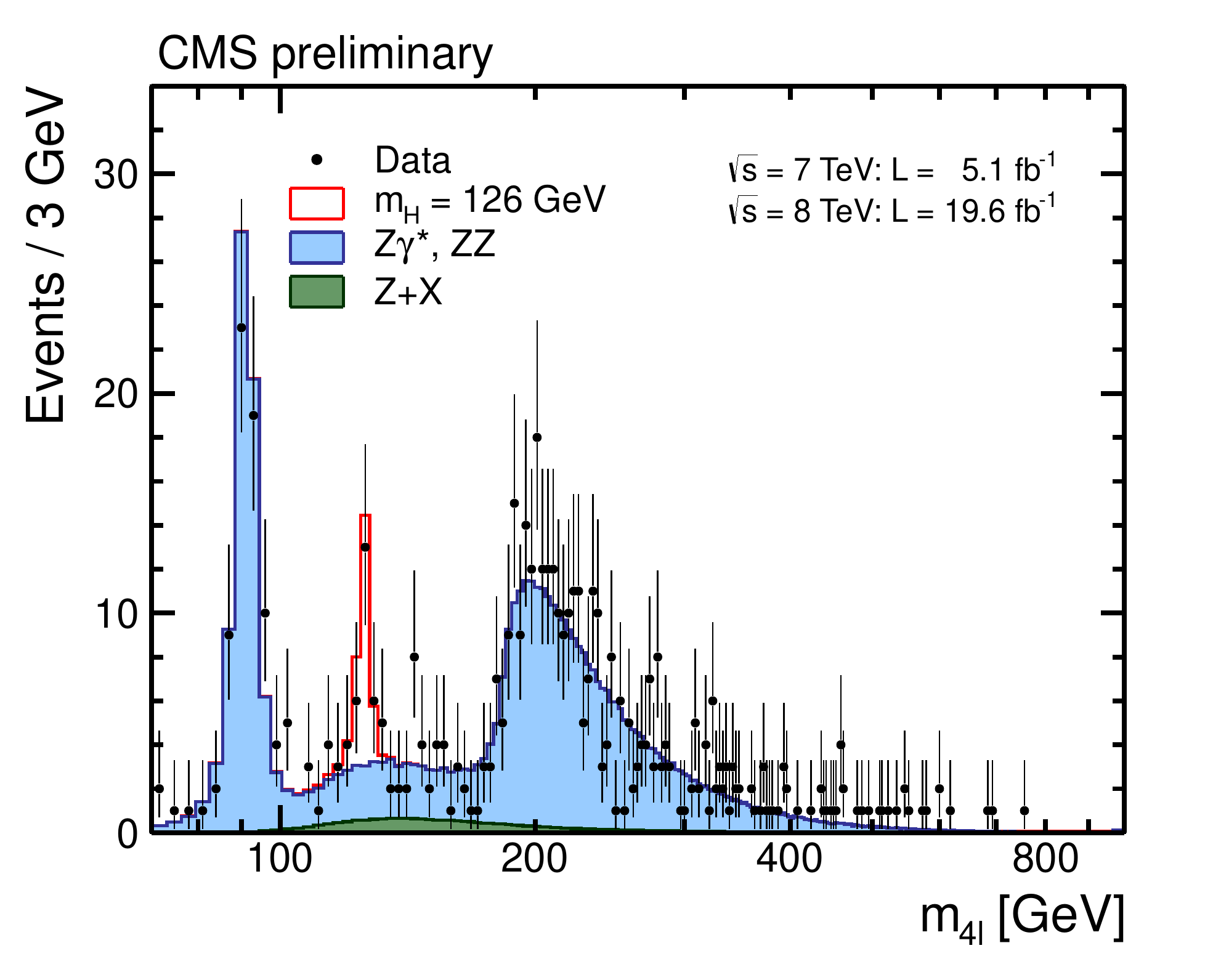}
\includegraphics[width=0.49\textwidth,height=0.25\textheight]{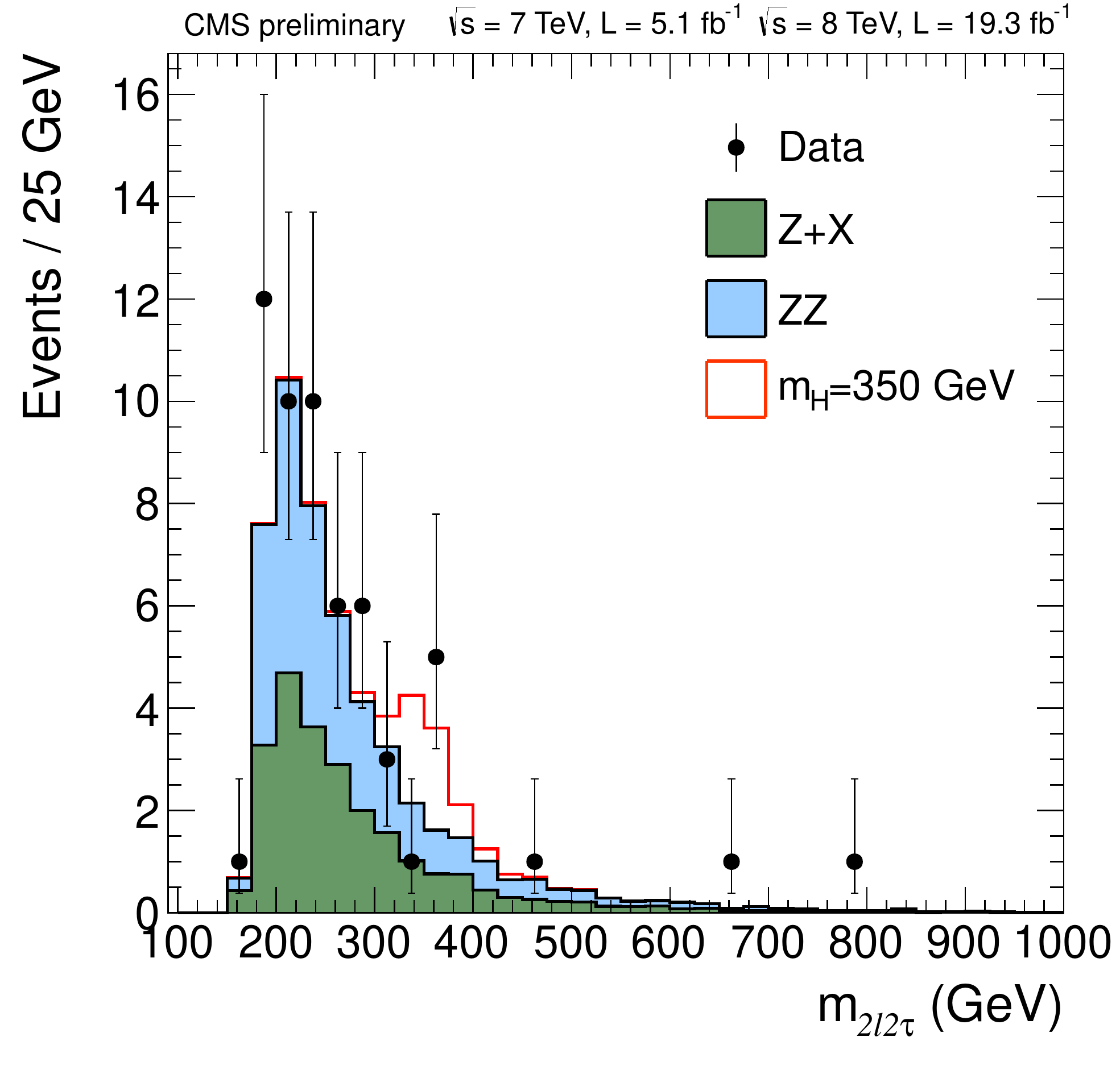}
\end{tabular}
\caption{ 
Distribution of the $4\ell$ (left) and $2\ell2\tau$ (right) reconstructed mass in the 
full mass range. Points represent the data, shaded histograms represent the background 
and unshaded histogram the signal expectation.}
\label{fig:Mass4l}
\end{center}
\end{figure}

The distributions of the kinematic discriminant $K_D$ versus $m_{4\ell}$ 
are shown for the selected events and compared to SM background expectation 
in Figure~\ref{fig:KDvsM4lFullMass}. The distributions of $\pt/m_{4\ell}$ and 
the VBF discriminant $V_D$ are presented in Figure~\ref{fig:VBF}.

\begin{figure}[!htb]
\vspace*{0.3cm}
\begin{center}
\centerline{
\includegraphics[width=0.49\linewidth,height=0.25\textheight]{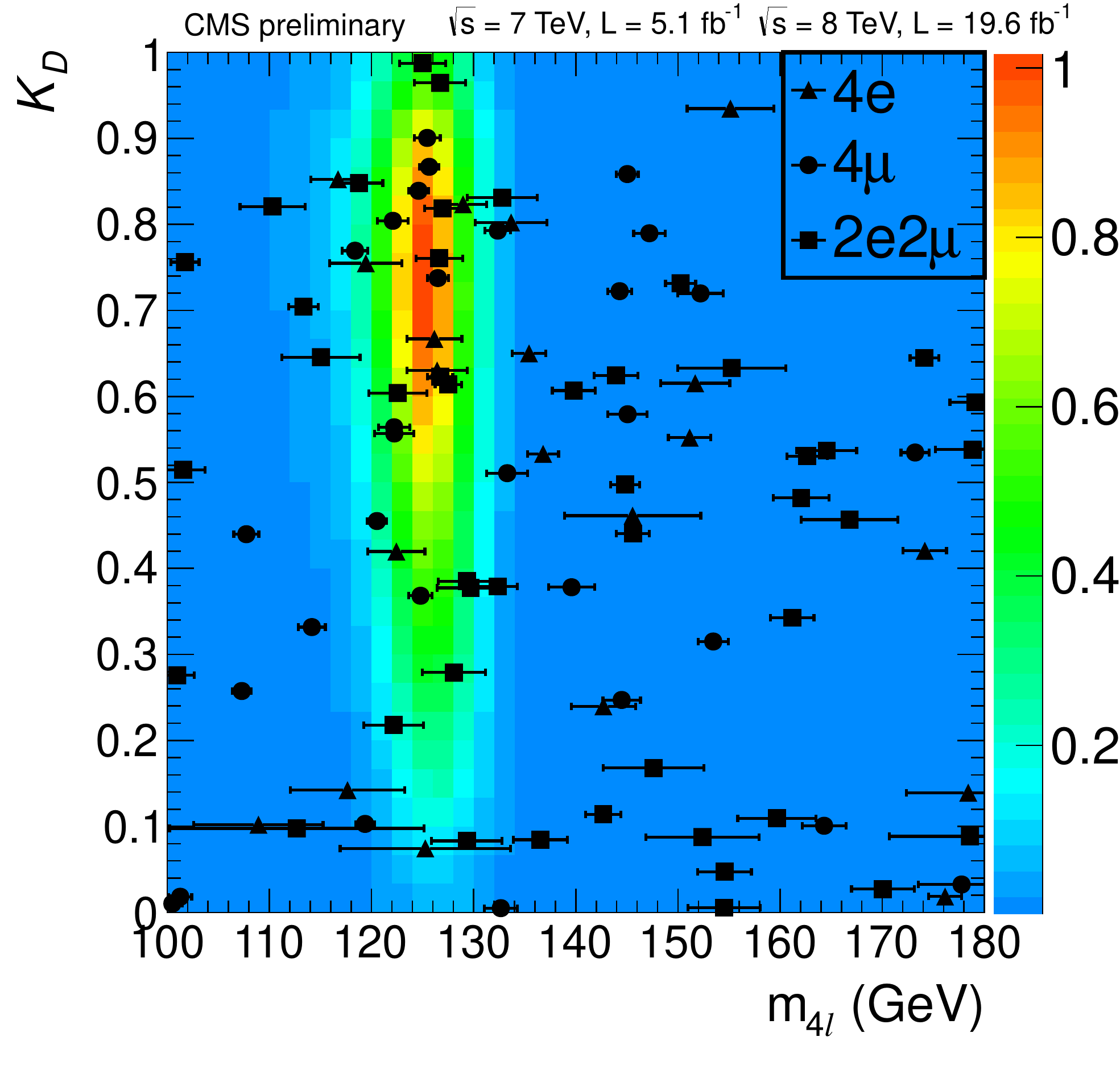}
\includegraphics[width=0.49\linewidth,height=0.25\textheight]{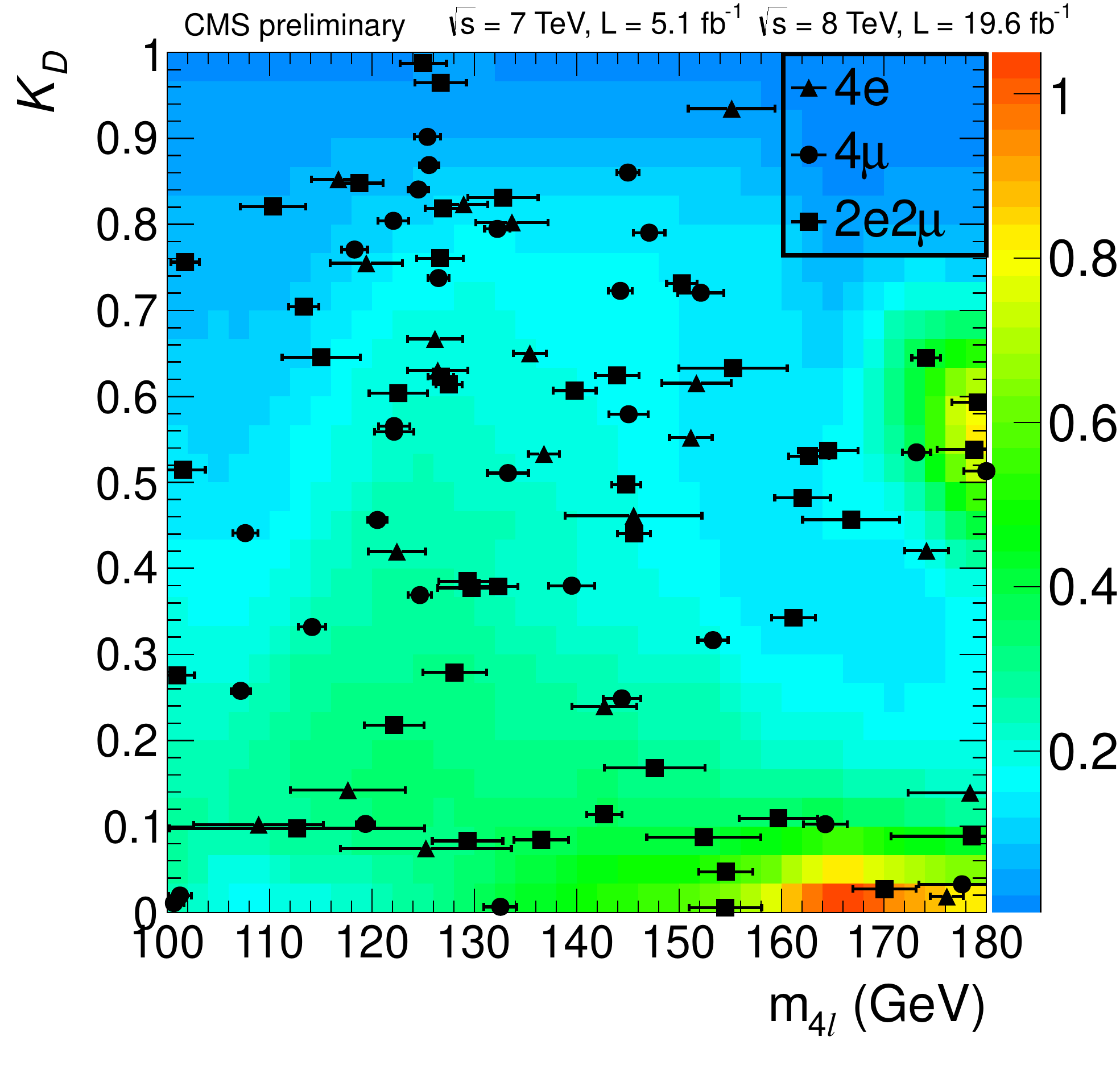}
}
\caption{ Distribution of the kinematic discriminant $K_D$ versus $m_{4\ell}$ in the low-mass region.
               The contours represent the expected relative density of signal events for $\mHi=126~\GeV$ (left) 
	       and for background events (right).
               The points show data and measured invariant mass uncertainties as horizontal bars.
}
\label{fig:KDvsM4lFullMass}
\end{center}
\end{figure}

%=============
\begin{figure}[!htb]
\vspace*{0.3cm}
\begin{center}
\begin{tabular}{cc}
\includegraphics[width=0.49\textwidth,height=0.25\textheight]{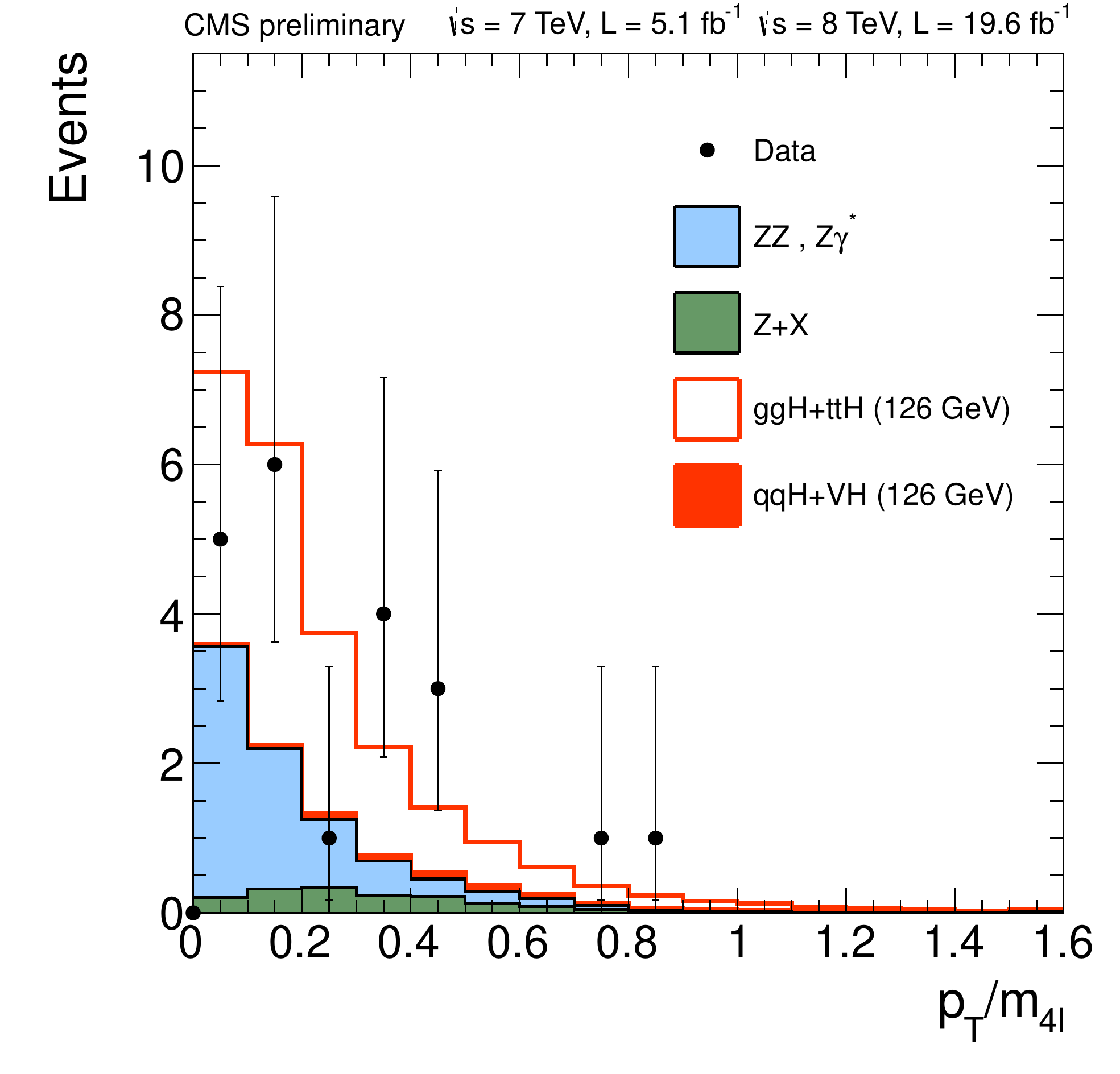}  
\includegraphics[width=0.49\textwidth,height=0.25\textheight]{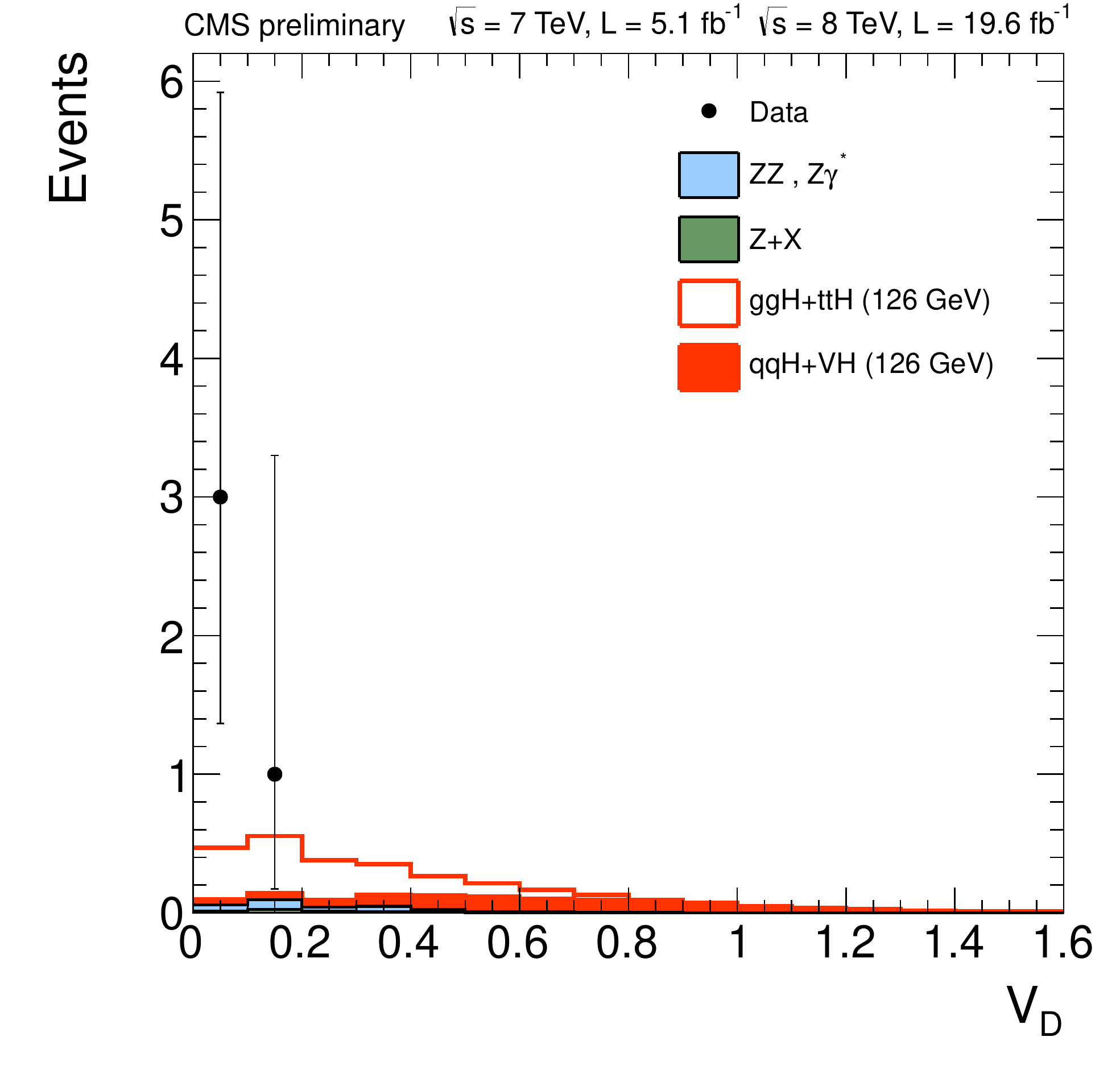}
%
% \vspace{-1.5cm}
\end{tabular}
\caption{ 
Distributions for the  $\pt / m_{4\ell}$ in the first category for the 
VBF discriminant in second category. Only events in the mass 
region $121.5 < m_{4\ell} < 130.5 \GeV$ are considered.
}
\label{fig:VBF}
\end{center}
\end{figure}
%=======

The local $p$-values, representing the significance of local excesses 
relative to the background expectation, are shown as a function 
of $\mHi$ in Figure~\ref{fig:UpperLimit_ASCLS-H}. The minimum 
of the local $p$-value is reached around $m_{4\ell} = 125.8$ GeV, and 
corresponds to a local significance of $6.7\sigma$ (for an expectation 
of 7.2 $\sigma$). This constitutes an observation of the new boson 
in the four-leptons channel alone. As a cross-check, the 1D ($m_{4\ell}$) 
and 2D ($m_{4\ell}$, $K_D$) models are also studied, 
and observed a local significance of 4.7 and 6.6 $\sigma$, for 
an expectation of 5.6 and 6.9 $\sigma$, respectively. 
The upper 95\% confidence level (CL) limits obtained from the combination of the $4\ell$ and $2\ell2\tau$ channels 
using the modified frequentist construction CL$_{s}$ 
method~\cite{LHC-HCG,Read1,junkcls} are also shown in Figure~\ref{fig:UpperLimit_ASCLS-H}. 
The SM-like Higgs boson is excluded by the four-lepton channels at 95\% CL 
in the range 130--827~$\GeV$ (for an expectation of 114--778 $\GeV$). 
The signal strength $\mu$, relative to the expectation for the SM Higgs boson,
is measured to be $\mu = 0.91^{+ 0.30}_{-0.24}$ at 125.8 $\GeV$. 

%%%%%%%%%%%%%%%%%%%%%%%
\begin{figure}[th!]
\begin{center}
\centerline{
\includegraphics[width=0.49\linewidth,height=0.25\textheight]{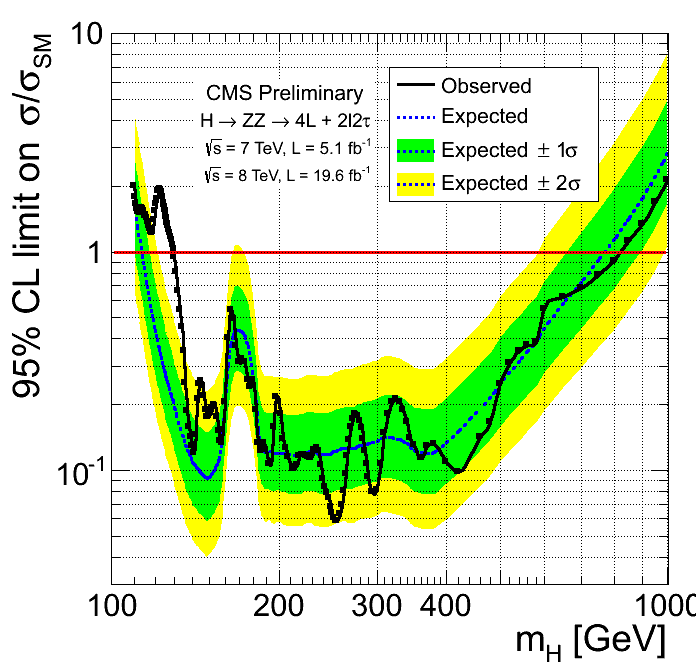}
\includegraphics[width=0.49\linewidth,height=0.25\textheight]{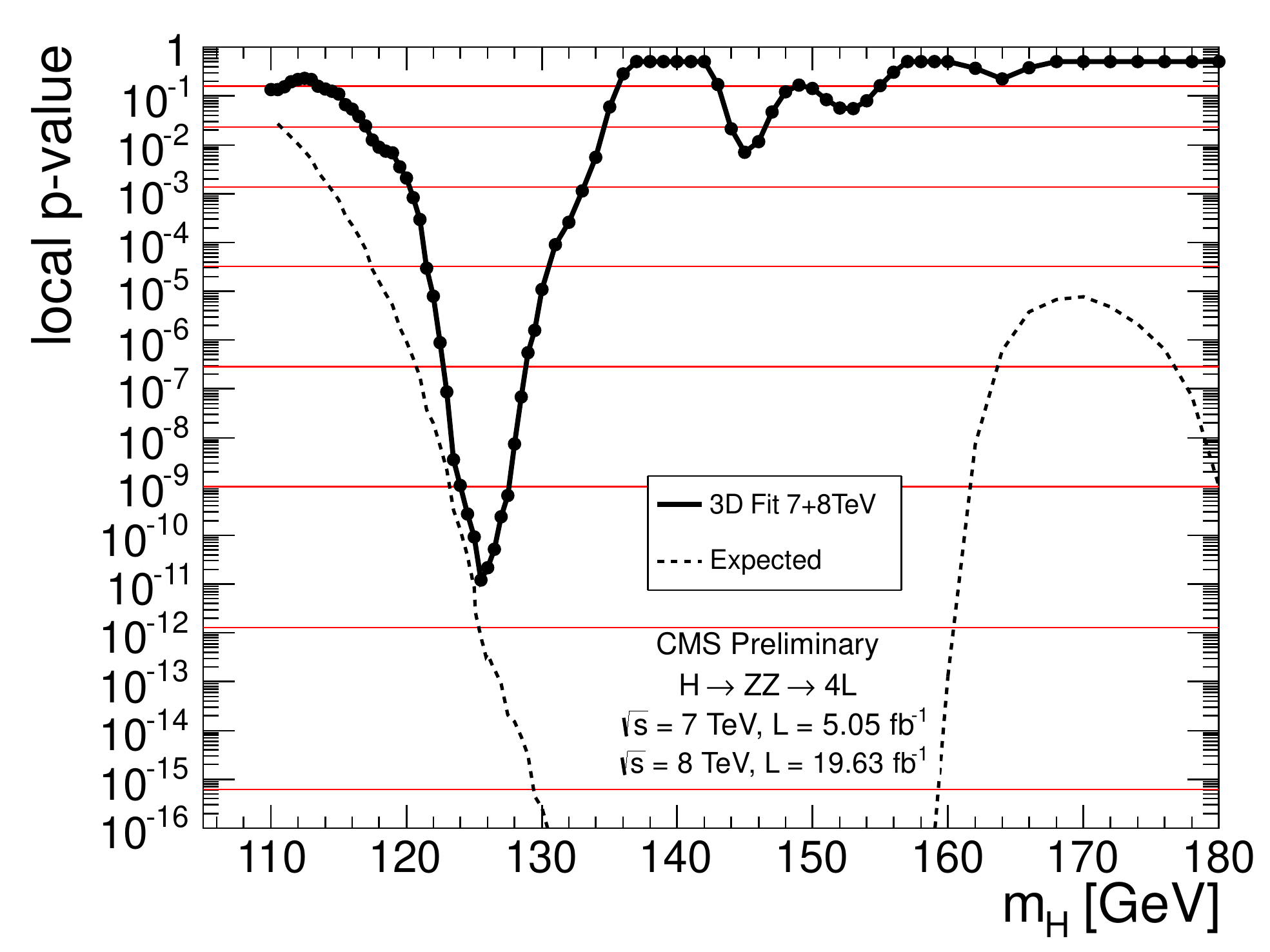}
}

\caption{
Observed and expected 95\% CL upper limit (left) on the ratio of the production cross section to the 
SM expectation in the $\Hi \to \Z\Z \to 4\ell$ analysis.
The 68\% and 95\% ranges of expectation  for the background-only model 
are also shown with green and yellow bands, respectively.
Significance of the local excess (right) with respect to the SM background 
expectation as a function of the Higgs boson mass.
}
\label{fig:UpperLimit_ASCLS-H}
\end{center}
\end{figure}
%%%%%%%%%%%%%%%%%%%%%%%

The mass measurement of the new resonance is performed with a three-dimensional fit using 
for each event the four-lepton invariant mass, the associated per-event mass uncertainty, and the 
kinematic discriminant. Per-event uncertainties on the four-lepton invariant mass are 
calculated from the individual lepton momentum uncertainties. Figure~\ref{fig:mass_measurement} 
shows the one-dimensional likelihood scan versus SM Higgs boson mass 
performed under the assumption that its width is much smaller than the detector 
resolution. The resulting fit gives
 $m_{\rm H} = 125.8 \pm 0.5$ (stat.) $\pm 0.2$ (syst.)~GeV. The systematic uncertainty 
 accounts for the effect on the mass scale of the lepton momentum scale and 
 resolution.
 
The jet categorization and the utilization of the transverse momentum 
spectrum and vector boson fusion sensitive variables are used to disentangle
the production mechanisms of the observed new state.
The production mechanisms are split into two categories depending on whether the production
is induced by vector bosons (VBF, VH) or fermions (gluon fusion loop with quarks, $\ttbar\Hi$).
Two respective signal strength modifiers ($\mu_F, \mu_V$)  are introduced 
as scale factors to the SM expected cross section. A two dimensional fit is performed
for the two signal strength modifiers assuming a mass hypothesis of $m_H = 125.8~\GeV$. 
Figure~\ref{fig:mass_measurement} shows the result of the ($\mu_V,\mu_F$) fit, 
leading to the measurements $\mu_V = 1.0^{+2.4}_{-2.3}$ and 
$\mu_F = 0.9^{+0.5}_{-0.4}$.

%=============
\begin{figure}[!htb]
\begin{center}
\includegraphics[width=0.49\textwidth,height=0.25\textheight]{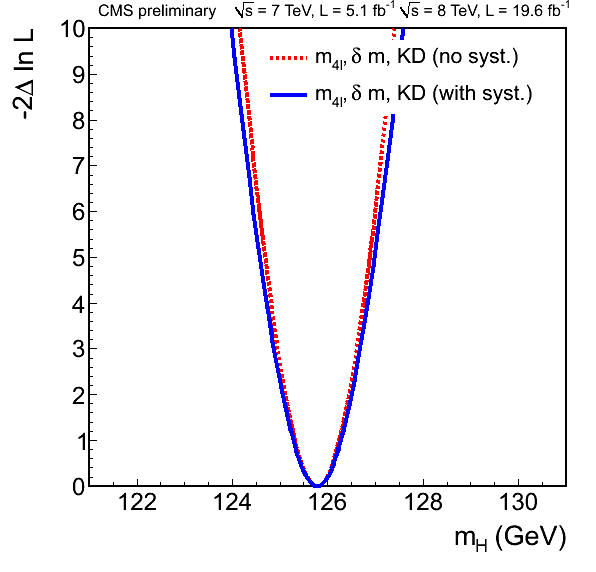}
\includegraphics[width=0.49\textwidth,height=0.25\textheight]{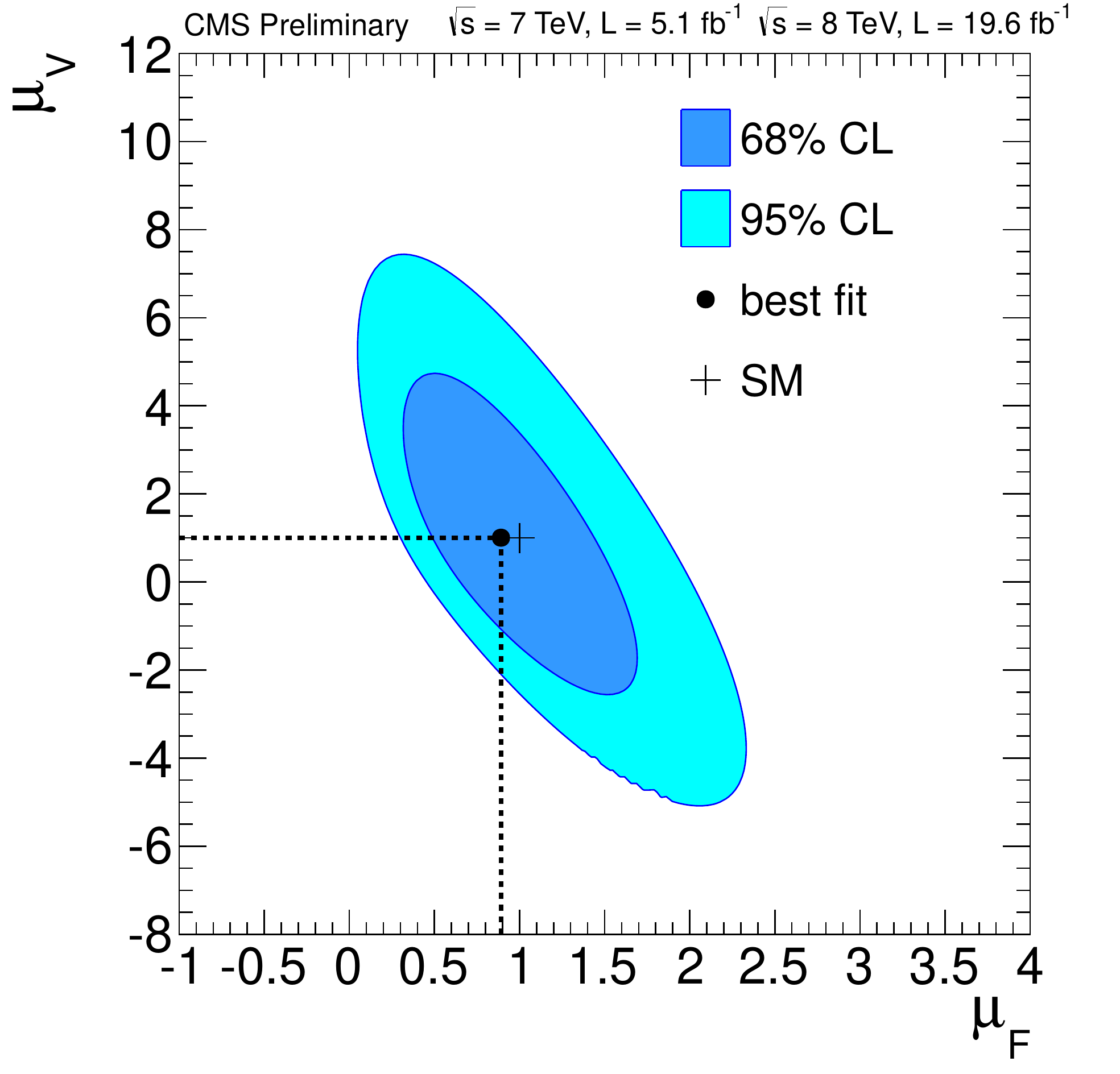}
\caption{
1D test statistics q($\mHi$)=$-2\Delta \ln {\rm L}$ scan vs tested Higgs boson mass~$\mHi$, obtained from 
the 3D test statistics profiling the minimum of the signal strengths, with and without systematics (left). 
Likelihood contours on  the signal strength modifiers associated with fermions 
($\mu_F$) and vector bosons ($\mu_V$) shown at 68\%  and 95\% CL (right).
}
\label{fig:mass_measurement}
\end{center}
\end{figure}
%%%%%%%%%%%%%%%%%%%%%%%%%%%%%%%%%%%%%%%%%%%%%%%%%%%%%%%%%%%%%%%%%%%%%%%%%%%%%%%
\section{$\Hi \to \W\W \to 2\ell 2\nu$ analysis}
\label{sec:hww}
%%%%%%%%%%%%%%%%%%%%%%%%%%%%%%%%%%%%%%%%%%%%%%%%%%%%%%%%%%%%%%%%%%%%%%%%%%%%%%%
The search strategy for $\hww$ is based on the final state in which
both $\W$ bosons decay leptonically~\cite{CMS-PAS-HIG-13-003}, resulting in a signature with two 
isolated, oppositely charged, high $\pt$ leptons (electrons or muons) 
and large missing transverse momentum, $\met$, due to the 
undetected neutrinos.

To improve the signal sensitivity, the events are separated according to lepton flavor
into $\Elp\Elm$, $\Mp\Mm$, and $\Elpm\Mmp$ samples and according to jet multiplicity into
0-jet and 1-jet samples.

To reduce the background from $\WZ$ production, any event
that has a third lepton passing the identification and isolation requirements is rejected.
The contribution from $\wgamma$ production, when the photon is misidentified as an 
electron, is reduced by about 90\% in the dielectron final state by $\GAMMA$ conversion 
rejection requirements. The background from low mass resonances is rejected by 
requiring a dilepton mass ($\mll$) greater 
than 12 $\GeV$. A minimum requirement on the dilepton transverse 
momentum ($\pt^{\ell\ell}$) is applied to reduce the $\Wjets$ background.

The Drell-Yan process produces same-flavor lepton pairs ($\Elp\Elm$ and $\Mp\Mm$). In order 
to suppress this background, a few additional cuts are applied in the same-flavor final states.
First, the resonant component of the Drell-Yan production is rejected by requiring
a dilepton mass outside a 30 $\GeV$ window centered on the $\Z$ pole.
Then, the remaining off-peak contribution is suppressed by exploiting different \met-based approaches. 

To enhance the sensitivity to a Higgs boson signal, a cut-based approach is chosen 
for the final ``Higgs" selection in all categories. Because the kinematics of 
signal events change as a function of the Higgs mass, separate optimizations are 
performed for different $\mHi$ hypotheses in a cut--based analysis. 
In addition, a two-dimensional shape 
analysis technique is also pursued for the different-flavor final state in 
the 0-jet and 1-jet categories. This second analysis is more sensitive to the 
presence of a Higgs boson and is used as a baseline for the final results.

In the cut-based approach extra requirements, designed to optimize the
sensitivity for a SM Higgs boson, are placed
on $\ptlmax$, $\ptlmin$, $\mll$, $\delphill$ and
the transverse mass $m_\mathrm{T}$,
defined as $\sqrt{2 \pt^{\ell\ell} \met (1-\cos\delphimetll)}$, where $\delphimetll$
is the difference in azimuth between $\met$ and the transverse momentum of the
dilepton system. The $\mll$ and $m_\mathrm{T}$ distributions 
in the 0-jet in the different-flavor final state are shown 
in Figure~\ref{fig:hwwsel_0j_mh125} for a SM Higgs boson with $\mHi=125~\GeV$.

\begin{figure}[h!]
\begin{center}
   \includegraphics[width=0.49\textwidth,height=0.25\textheight]{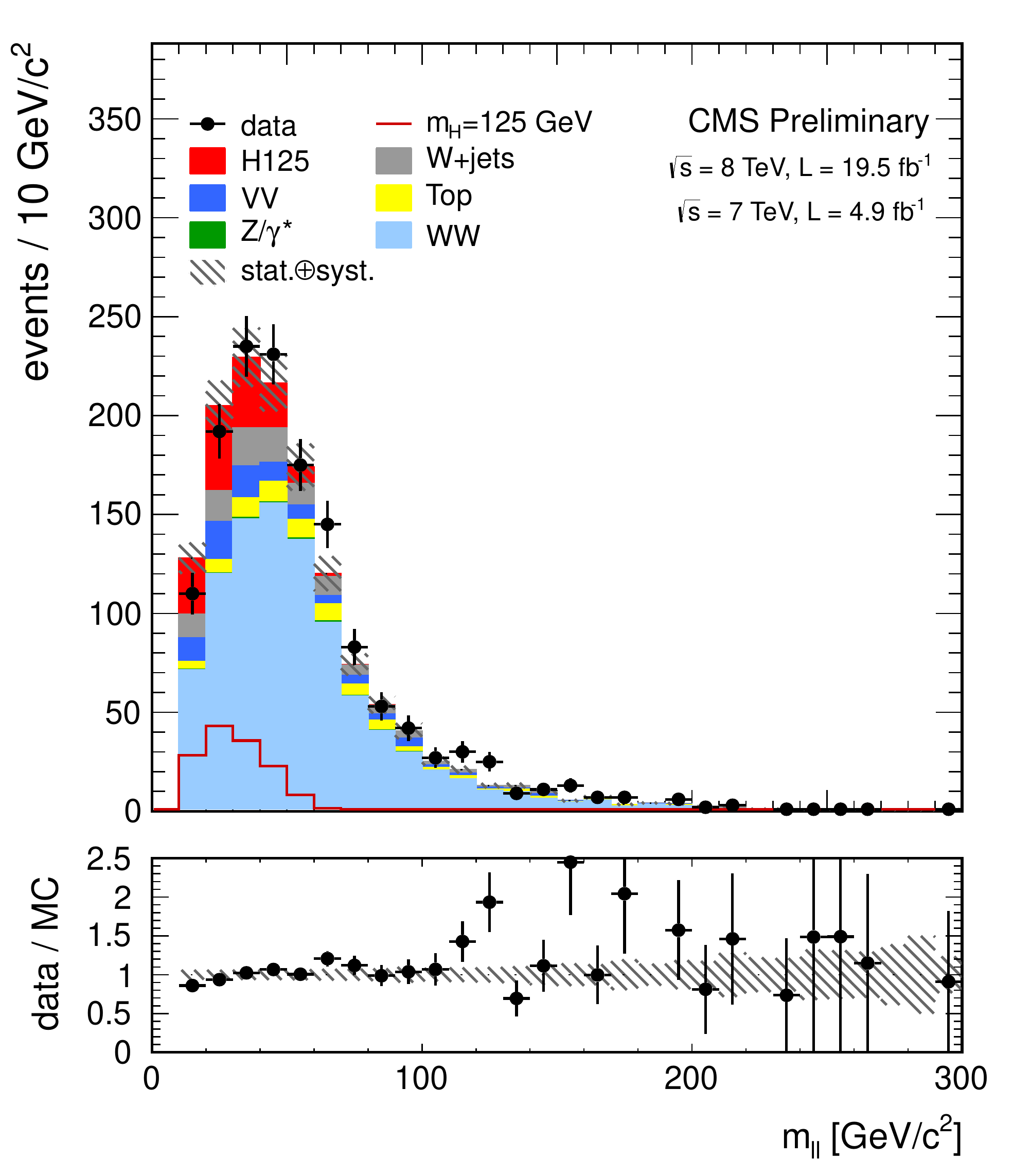}
   \includegraphics[width=0.49\textwidth,height=0.25\textheight]{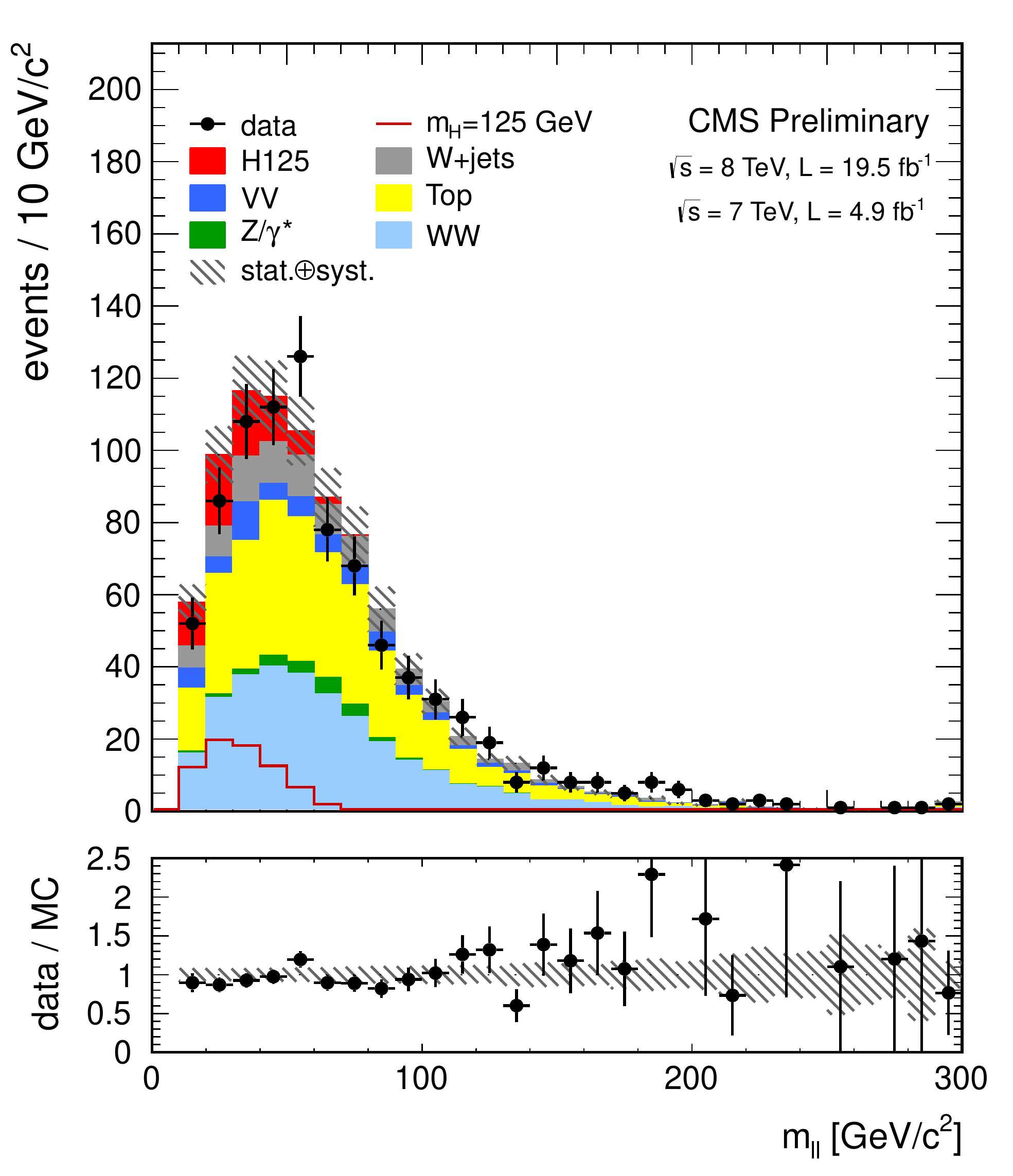}
	\caption{Distributions of dilepton mass (left) and the transverse mass (right) in the 
	0-jet category, in the different-flavor final state for a $\mHi=125~\GeV$ SM Higgs boson 
	and for the main backgrounds.
       The cut-based $\hww$ selection, except for the requirement on the variable itself, 
       is applied.}  \label{fig:hwwsel_0j_mh125}
\end{center}
\end{figure}

The two-dimensional shape analysis for the different-flavor final state 
uses two independent variables, $m_\mathrm{T}$ and $\mll$. It allows for a 
simpler physical interpretation of the observed data with 
a sensitivity comparable to other more complex techniques.
The two-dimensional distributions for the $\mHi = 125~\GeV$ Higgs signal hypothesis and 
background processes are shown in Figure~\ref{fig:histo_2D_0j} for the 0-jet bin.

\begin{figure*}[htb]
\begin{center}
 \includegraphics[width=0.49\textwidth,height=0.25\textheight]{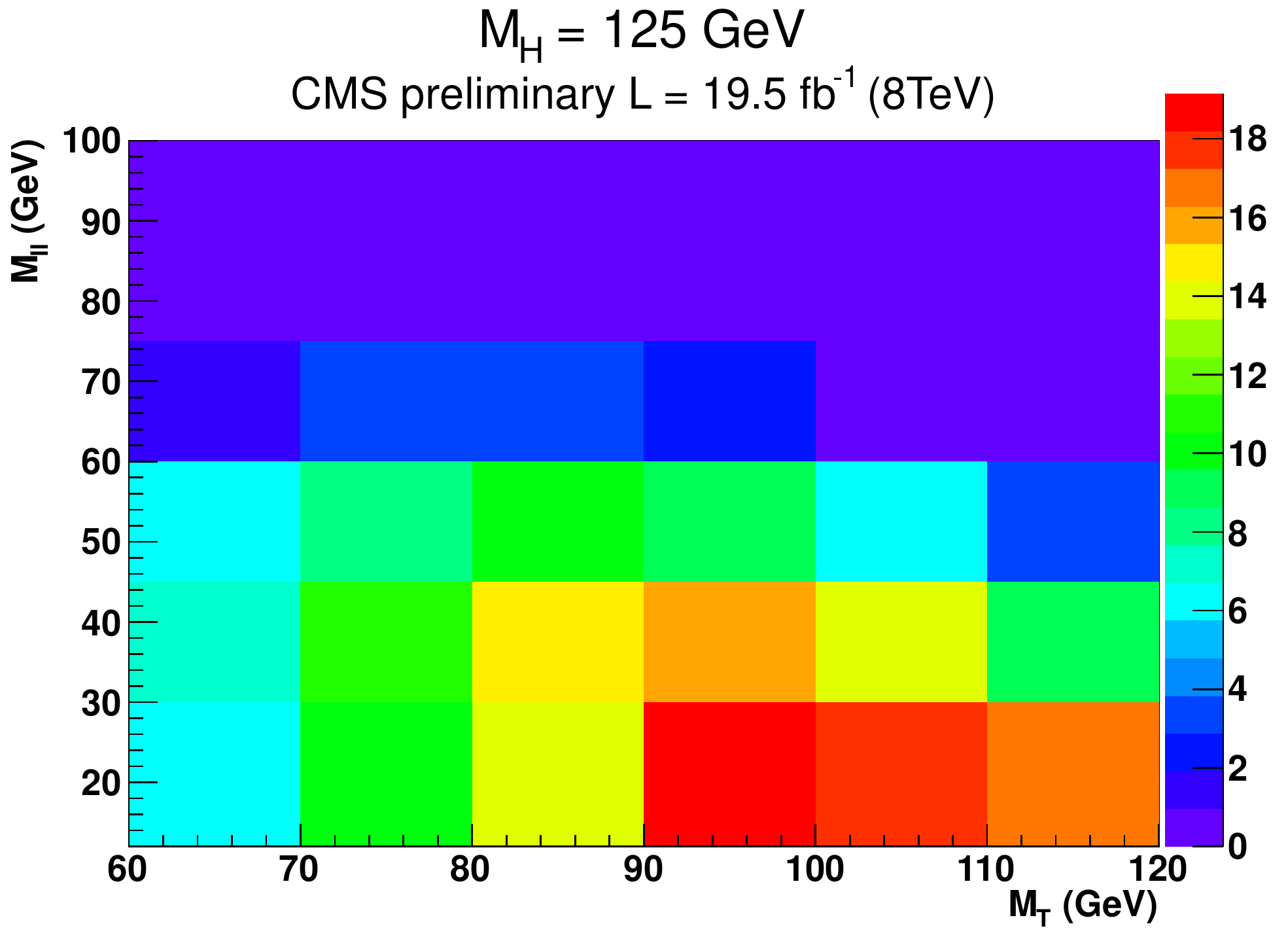}
 \includegraphics[width=0.49\textwidth,height=0.25\textheight]{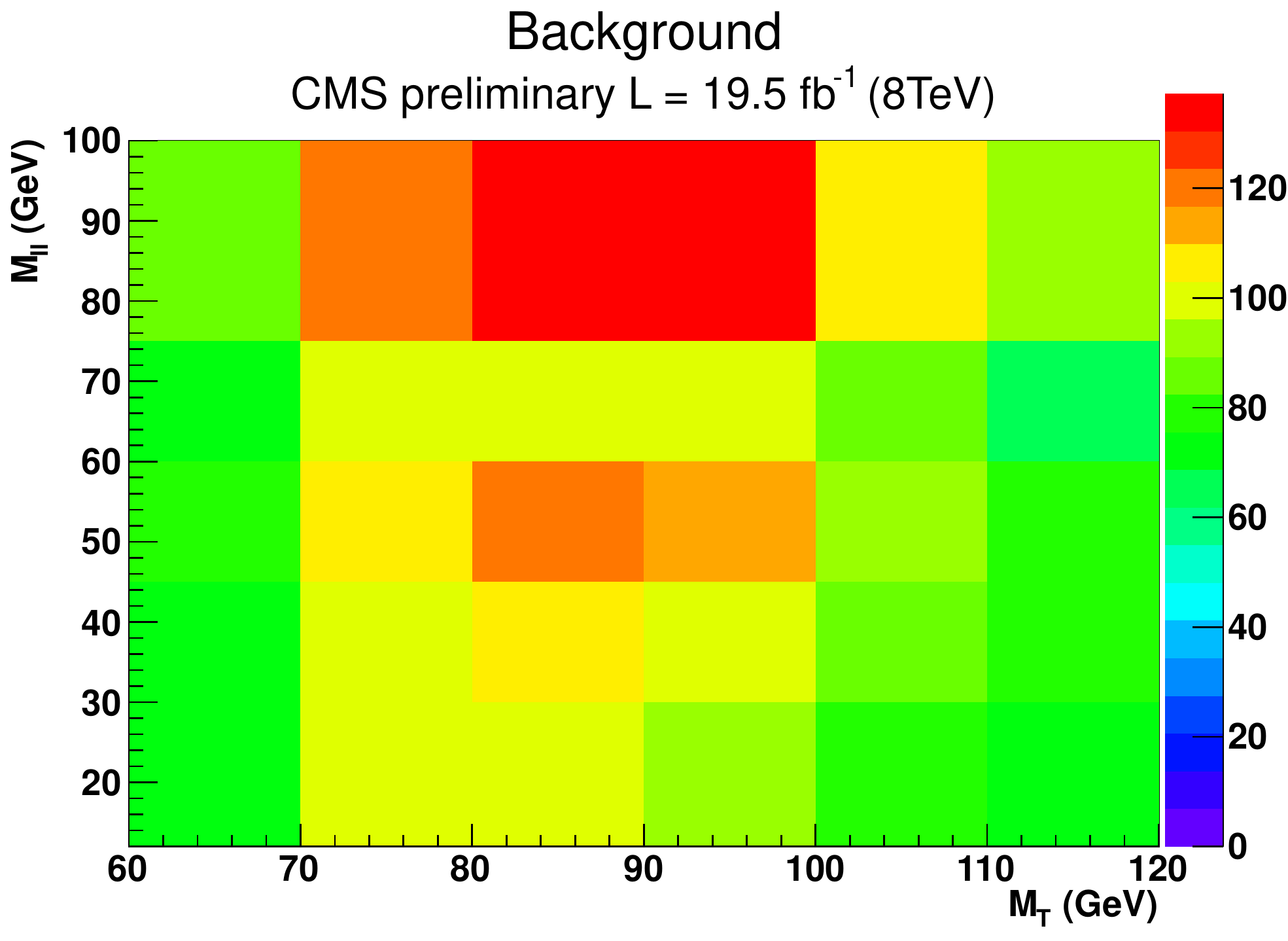}
       \caption{Two-dimensional $\mt-\mll$ distributions in the 0-jet bin for 
       the $\mHi = 125~\GeV$ SM Higgs signal hypothesis (left) and the background processes (right).}
\label{fig:histo_2D_0j}
\end{center}
\end{figure*}

After applying the Higgs selection, upper limits 
are derived for the ratio of the product of the Higgs boson 
production cross section and the $\Hi \to \WW$ branching fraction,
$\sigma_{\Hi} \times \mathrm{BR}(\Hi \to \WW)$, and the
SM Higgs expectation, $\sigma/\sigma_{SM}$. For the shape approach, the analysis 
in the different-flavor final state in the 0-jet and 1-jet categories is combined with 
the cut-based analysis in all other categories. 

The 95\% observed and median expected CL upper limits for the shape analysis 
are shown in Figure~\ref{fig:xsLimCuts}, which excludes a Higgs boson 
in the mass in the range 128--600 $\GeV$ at 95\% CL. 
The expected exclusion range for the background only hypothesis is 115--575 $\GeV$. 
An excess of events is observed for hypothetical 
low Higgs boson masses, which makes the observed limits weaker than the 
expected ones. Due to the poor mass resolution of this channel the excess
extends over a large mass range.

The observed (expected) significance for a SM Higgs with a mass of 125 $\GeV$ 
is 4.0 (5.1) standard deviations for the shape-based analysis. 
The observed and expected significances and for each Higgs mass hypothesis 
is shown in Figure~\ref{fig:xsLimCuts}. 
The observed $\mu$ value for $\mHi = 125~\GeV$ using the shape-based analysis is 
0.76 $\pm$ 0.13 (stat.) $\pm$ 0.16 (syst.) = 0.76 $\pm$ 0.21 (stat.+syst.). The
statistical component is obtained by fixing all the nuisance parameters to their
fit values and recomputing the likelihood profile. Then, the systematic
component comes from a subtraction in quadrature of the full uncertainty and the
statistical component.

\begin{figure}[htbp]
  \begin{center}
  \includegraphics[width=0.49\textwidth,height=0.25\textheight]{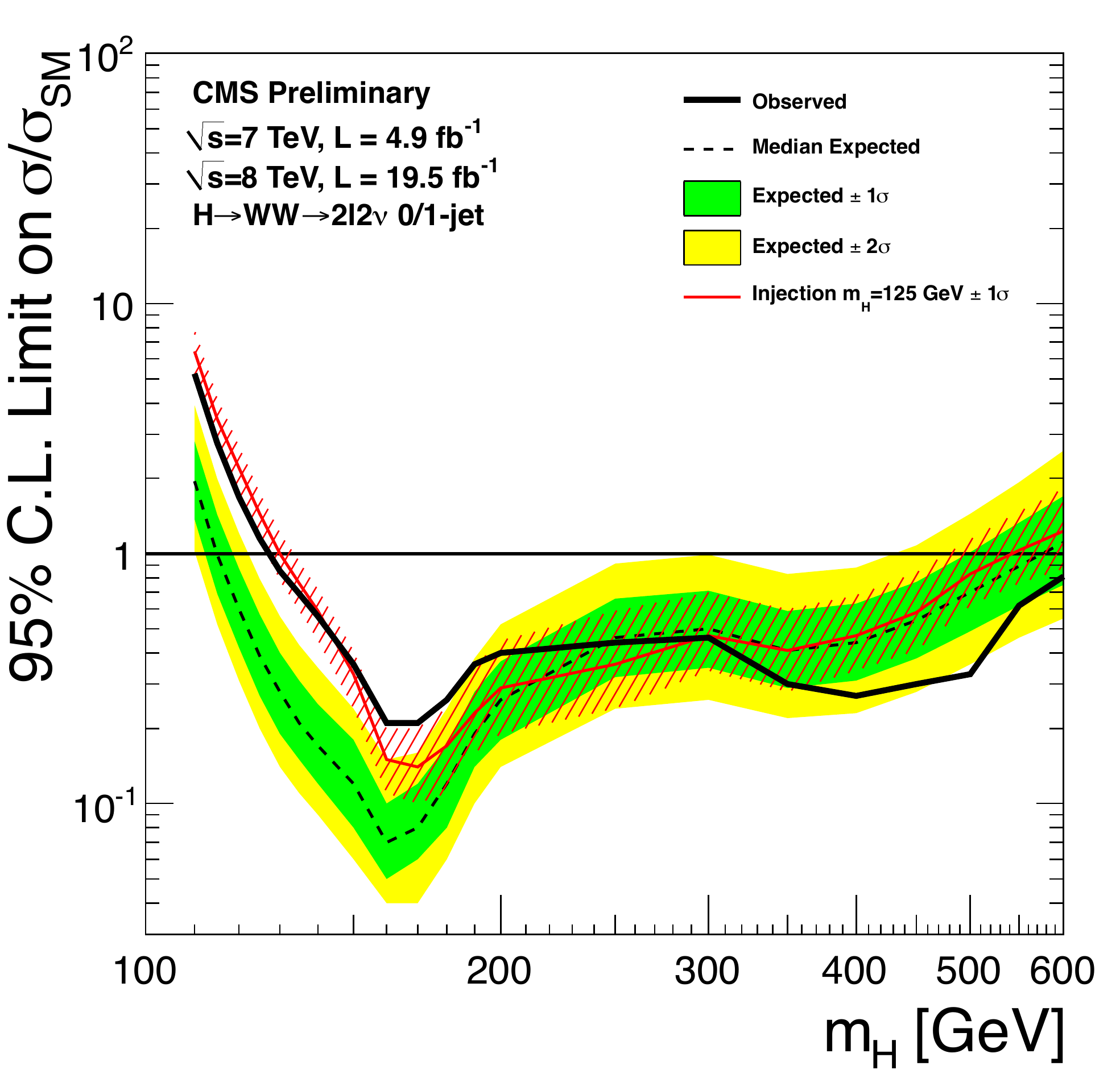}
  \includegraphics[width=0.49\textwidth,height=0.25\textheight]{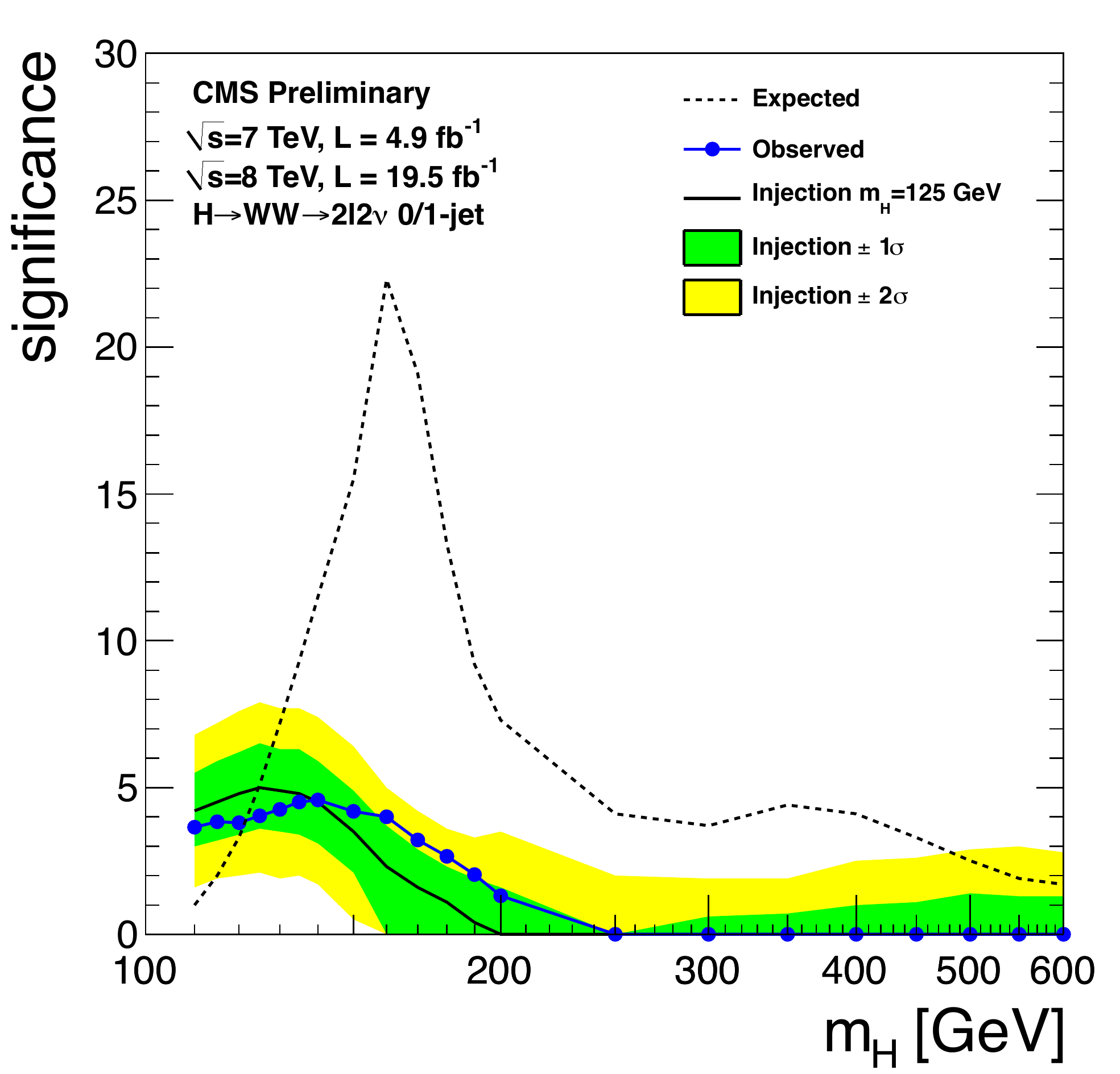}
    \caption{Expected and observed 95\% CL upper
       limits on the cross section times branching fraction,
       $\sigma_{\Hi} \times \mathrm{BR}(\Hi \to \WW)$,
       relative to the SM Higgs expectation for the shape-based analysis (left).
       The expected limits in the presence of the Higgs 
       with $\mHi = 125 \GeV$ and its associated uncertainty are also shown. 
       The observed and expected significances and for each Higgs mass hypothesis (right) for the shape-based 
       analysis. The expected significance under the presence of a $\mHi = 125 \GeV$ Higgs is also 
       shown.}
    \label{fig:xsLimCuts}
  \end{center}
\end{figure}

%%%%%%%%%%%%%%%%%%%%%%%%%%%%%%%%%%%%%%%%%%%%%%%%%%%%%%%%%%%%%%%%%%%%%%%%%%%%%%%
\section{$\W\Hi \to \W\W\W \to 3\ell 3\nu$ analysis}
\label{sec:wh}
%%%%%%%%%%%%%%%%%%%%%%%%%%%%%%%%%%%%%%%%%%%%%%%%%%%%%%%%%%%%%%%%%%%%%%%%%%%%%%%
In the $\W\Hi \to \W\W\W \to 3\ell 3\nu$ channel~\cite{CMS-PAS-HIG-13-009}, 
events are selected by requiring three charged 
lepton candidates, electrons or muons, with total charge equal to $\pm$1 are 
required, with $\pt >$20 $\GeV$ for the leading lepton and 
$\pt >$10 $\GeV$ for the other leptons. The data analysis is performed by 
using a shape-based approach, with a cross check from a single bin counting 
experiment. To further improve the sensitivity the events are split into two categories: 
all events that have an opposite-sign same-flavor lepton pair are classified in one category 
(OSSF), everything else is classified as in the same-sign same-flavor category
(SSSF). While 1/4 of the events are selected in the second category, the 
expected background is rather small since physics processes leading to
this final state have small cross section.

Events are required to have $\met$ above 40 (30) $\GeV$ in the OSSF (SSSF) 
category. To reduce the background from top decays, events are rejected 
if there is at least one jet with $\Et$ above 40 $\GeV$. 
The $\WZ \to 3\ell\nu$ background is largely reduced by requiring that all 
the OSSF lepton pairs have a dilepton mass at least 25 $\GeV$ away from $m_{\Z}$. 
To reject the $\Z/\W+\gamma^{*}$ background, the dilepton mass of all
opposite-charge lepton pairs are required to be greater than 12 $\GeV$. 
Finally, the signal region is defined by requiring in addition to all the 
above cuts that the smallest dilepton mass $\mll$ is less than 100 $\GeV$ 
and that the smallest distance between opposite-charge leptons 
$\Delta R_{\ell^+\ell^-}$  is less than 2.

A shape-based analysis is carried out as a main analysis due to its superior 
performance with respect to a simple counting experiment. In this analysis a cut on
$\Delta R_{\ell^+\ell^-}$ is not applied, and instead it is used as the final 
discriminant. Tests have shown this variable to provide the best discrimination 
between signal and background events, in terms of both expected limits and significance.

No significant excess of events is observed with respect to the background 
prediction, and 95\% CL upper limits are calculated for the 
Higgs boson cross section with respect to $\sigma/\sigma_{SM}$. 
The expected and observed upper limits are shown in Figure~\ref{fig:combined_wh3l_from110to200_logx0_logy1}. 
Since the analysis is independent of Higgs mass, and the shape of the 
$\Delta R_{\ell^+\ell^-}$ distribution changes just slightly for the Higgs signal, 
only small fluctuations are expected between different Higgs mass hypotheses. 
For the cut-based analysis, the observed (expected) upper limit at the 95\% CL is 3.7 (3.6) 
times larger than the SM expectation for $\mHi=125~\GeV$. 
For the shape-based analysis, the observed (expected) upper limit at the 95\% CL is 3.3 (3.0)
times larger than the SM expectation for $\mHi=125~\GeV$.

\begin{figure}[htbp!]
\begin{center}
   \includegraphics[width=0.49\textwidth,height=0.25\textheight]{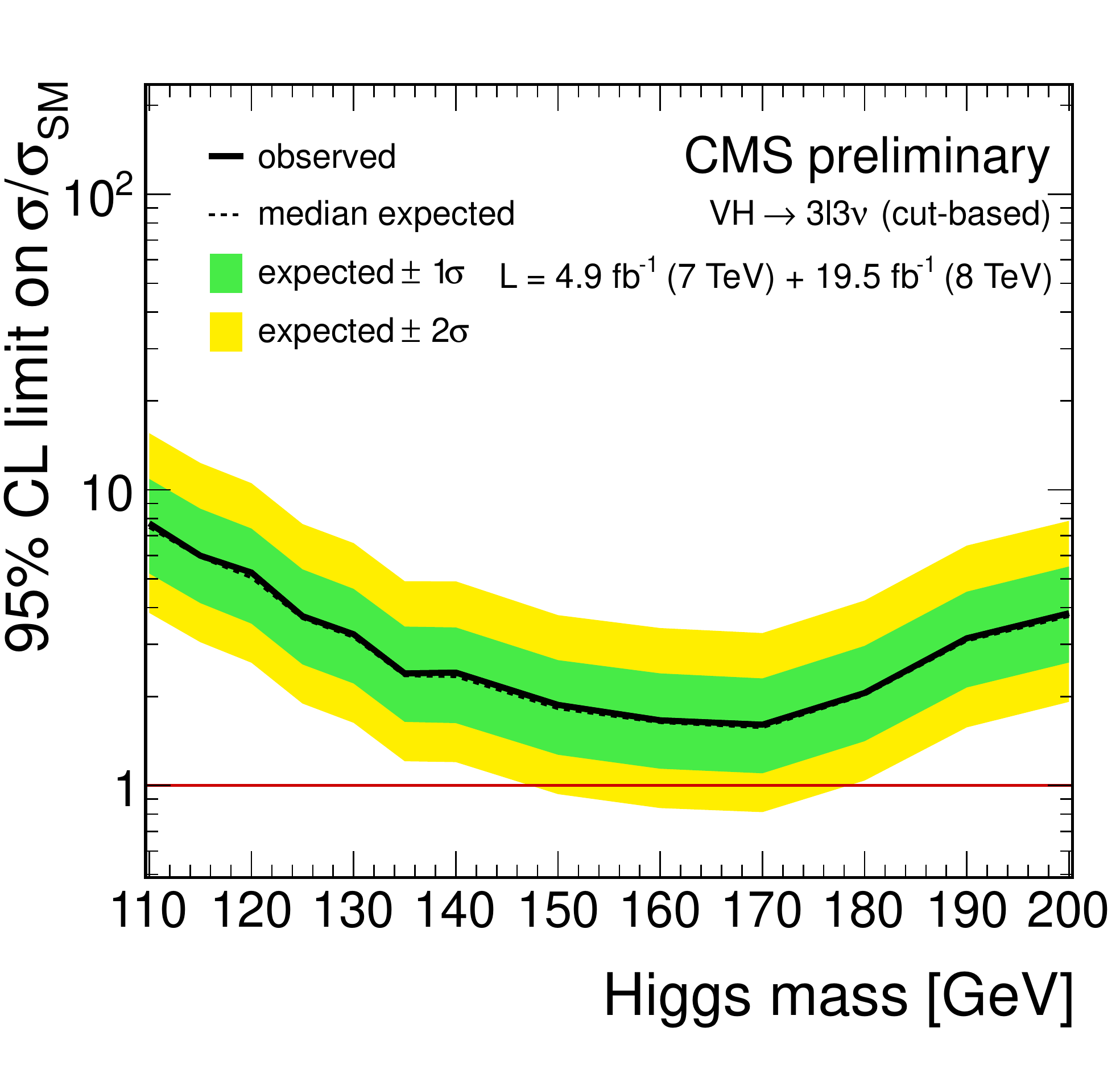}
   \includegraphics[width=0.49\textwidth,height=0.25\textheight]{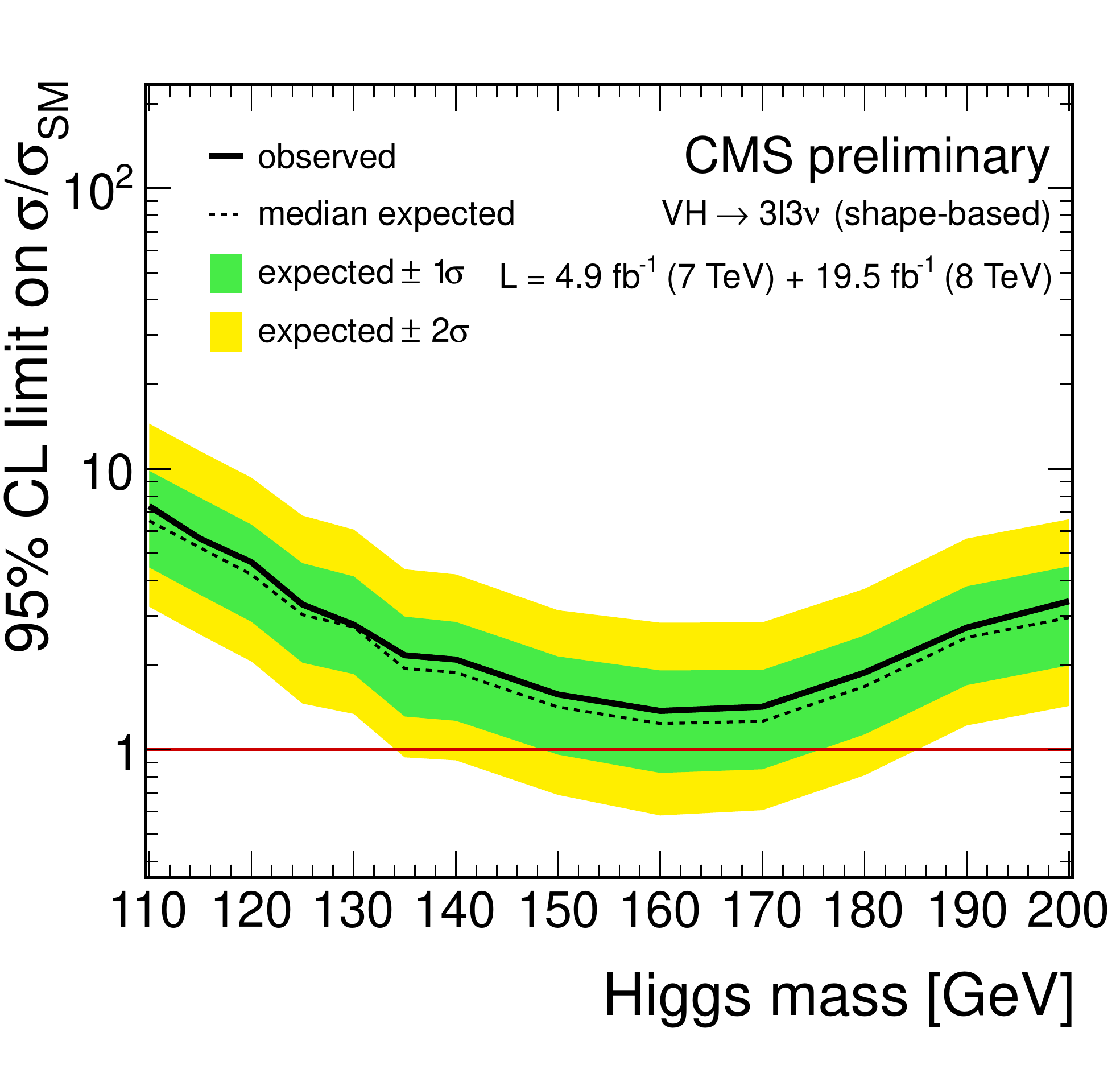}
   \caption{Upper limits at 95\% CL in the $\W\Hi \to 3\ell 3\nu$ final state for the cut-based 
   analysis (left) and shape-based analysis (right)}
   \label{fig:combined_wh3l_from110to200_logx0_logy1}
\end{center}
\end{figure}

%%%%%%%%%%%%%%%%%%%%%%%%%%%%%%%%%%%%%%%%%%%%%%%%%%%%%%%%%%%%%%%%%%%%%%%%%%%%%%%
\section{$\Hi \to \Z\gamma$ analysis}
\label{sec:hzg}
%%%%%%%%%%%%%%%%%%%%%%%%%%%%%%%%%%%%%%%%%%%%%%%%%%%%%%%%%%%%%%%%%%%%%%%%%%%%%%%
The $\Hi \to \Z\gamma$ decay channel~\cite{CMS-PAS-HIG-13-006} is a clean final state, with the $\Z$ boson
decaying into an electron or a muon pair, plus an isolated photon.

The invariant mass of at least one $\ell^+\ell^-$  pair is required to be greater 
than 50 $\GeV$. 
If two dilepton pairs are present, the one closest to the $\Z$ mass is taken.
The invariant mass of the $\ell^+\ell^-\gamma$ system, $m_{\ell\ell\gamma}$, is required to 
be between 100 and 180\,\GeV. Other conditions that combine the information from the photon and the leptons are:
(1) the ratio of the photon transverse energy to $m_{\ell\ell\gamma}$ must be greater than 15/110,
this requirement allows us 
to reject backgrounds without significant loss in signal sensitivity and without introducing a bias
in the $m_{\ell\ell\gamma}$ spectrum;
(2) the $\Delta R$ separation between each lepton and the photon must be greater than
0.4~in order to reject events with initial-state radiation avoiding photon influence lepton isolation;
and (3) final-state radiation events are rejected by requiring a  minimum
of 185\,GeV on the sum of $m_{\ell\ell\gamma}$ and $m_{\ell\ell}$.

The sensitivity of the search is enhanced by subdividing the selected events into classes
according to indicators of the expected mass resolution and the signal-to-background ratio,
and then  combine the results in each class.
For this purpose, four mutually exclusive event classes are defined: in terms of the
pseudo-rapidity of the leptons and the photon and on the shower shape
 of the photon for one of the topologies. The background model fit to 
the $m_{\mu\mu\gamma}$ distribution for two event classes is shown in Figure~\ref{fig:hzg_limits}.

No excess over the background is observed, and therefore
the data are used to derive upper limits on the proton-proton  Higgs boson production cross
section times the  $\Hi\to\Z\gamma$  branching fraction, $\sigma_{\Hi} \times \mathrm{BR}(\Hi \to \Z\gamma)$. 
The expected and observed limits are both shown in Figure~\ref{fig:hzg_limits}. 
The expected exclusion limits at 95\% confidence level are between 6 and 19 times 
the standard model cross section and the observed limit fluctuates between about 
3 and 31 times the SM cross section.  

\begin{figure}[hbtpH]
  \begin{center}
    \includegraphics[width=0.49\textwidth,height=0.25\textheight]{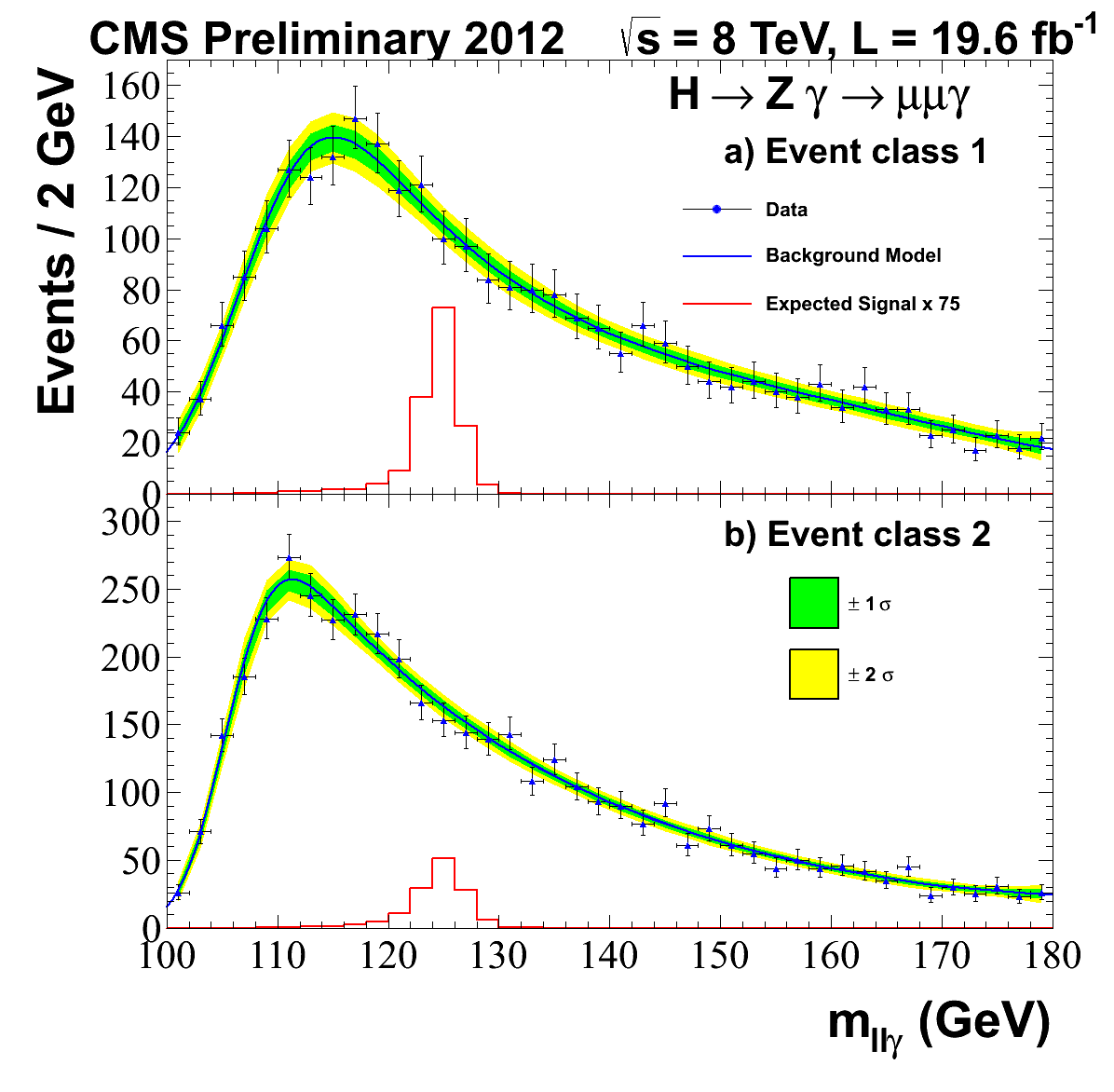}
    \includegraphics[width=0.49\textwidth,height=0.25\textheight]{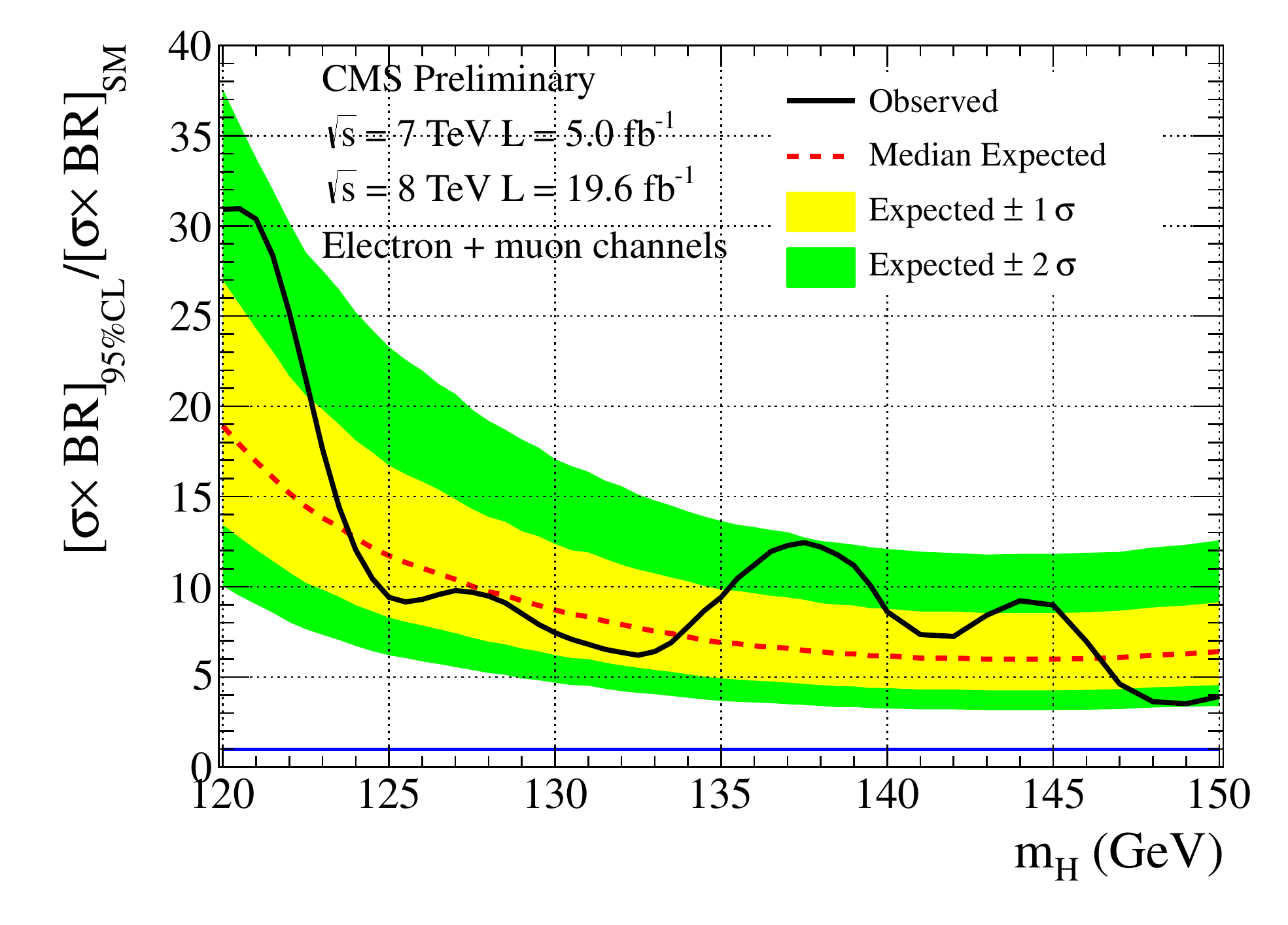}
    \caption{Background model fit to the $m_{\mu\mu\gamma}$ distribution for two event classes (left). 
    The statistical uncertainty bands shown are computed from the data fit. Exclusion limit on the 
    cross section of a SM Higgs boson decaying into \Z-boson and a photon as a function of $\mHi$ (right).}
    \label{fig:hzg_limits}
  \end{center}
\end{figure}

\section{Summary}
The status of the SM Scalar Boson search in the bosonic decay channels at the 
CMS experiment at the LHC has been presented. The results are based on proton-proton collisions 
data corresponding to integrated luminosities of up to 5.1 $\ifb$ at $\sqrt{s}$ = 7 $\TeV$ 
and 19.6 $\ifb$ at $\sqrt{s}$ = 8 $\TeV$. The observation of a new boson at a mass near 
126 $\GeV$ is confirmed by the analysis of the new data and first measurements of the boson 
properties have been shown.

\end{document}

%%%%%%%%%%%%%%%%%%%%%%
% End of moriond.tex  %
%%%%%%%%%%%%%%%%%%%%%%

%%% Local Variables: 
%%% mode: latex
%%% TeX-master: t
%%% End: 

%%% Local Variables: 
%%% mode: latex
%%% TeX-master: t
%%% End: 

%%% Local Variables: 
%%% mode: latex
%%% TeX-master: t
%%% End: 